\documentclass[a4paper,11pt]{article}
\pdfoutput=1
\usepackage{jheppub}
\usepackage{amsthm,mathtools}
\usepackage{enumitem}
\usepackage{tikz}
\usetikzlibrary{arrows.meta,positioning}

\newtheoremstyle{jplain}{\topsep}{\topsep}{\itshape}{}{\scshape}{.}{.5em}{}
\newtheoremstyle{jdef}{\topsep}{\topsep}{\normalfont}{}{\scshape}{.}{.5em}{}
\theoremstyle{jplain}
\newtheorem{theorem}{Theorem}[section]
\newtheorem{proposition}[theorem]{Proposition}
\newtheorem{lemma}[theorem]{Lemma}
\newtheorem{corollary}[theorem]{Corollary}
\theoremstyle{jdef}

\numberwithin{equation}{section}

\newcommand{\R}{\mathbb{R}}
\newcommand{\Z}{\mathbb{Z}}
\newcommand{\C}{\mathbb{C}}
\newcommand{\Hb}{\mathbb{H}}
\newcommand{\SL}{\mathrm{SL}(2,\R)}
\newcommand{\PSL}{\mathrm{PSL}(2,\R)}
\newcommand{\tSL}{\widetilde{\mathrm{SL}}(2,\R)}
\newcommand{\Diff}{\mathrm{Diff}^+(S^1)}
\newcommand{\Vect}{\mathrm{Vect}(S^1)}
\newcommand{\Ind}{\mathrm{Ind}}
\newcommand{\tr}{\mathrm{tr}}

\newcommand{\Om}{\mathcal{O}}
\newcommand{\Sch}{\mathcal{S}}
\newcommand{\sltwo}{\mathfrak{sl}(2,\R)}

\title{Finite cyclic monodromy in BMS$_3$ gravity}

\author{Evangelos Melas}
\affiliation{Department of Mathematics, School of Science, University of Thessaly,\\ 35100 Lamia, Greece}
\emailAdd{evanmelas@uth.gr}

\abstract{We investigate inducing data for three dimensional asymptotically flat BMS gravity on hyperbolic Virasoro orbits with winding, which contain no constant supermomentum, and the restriction of formal smooth equivariant sections to global Lorentz orbits. Using the developing map of the associated Hill operator we rederive their stabilizer and show that, in the spin double cover, it is the direct product of a line with a cyclic group of even order. The character of the cyclic factor is therefore a discrete inducing label independent of the continuous one, unlike for the massive BMS particles, where it is fixed by the spin. At points with the corresponding finite rotational symmetry, restriction of formal smooth equivariant sections to a global Lorentz orbit yields the finite cyclic equivariance condition, while an independently defined unitary global branch realizes the associated finite difference equation. At points with timelike projected momentum we derive the resulting ladder of Poincar\'e helicities, show that the projected rest energy is not an orbit invariant, construct locally flat winding BMS geometries with hyperbolic Lorentz holonomy, prove that the smooth real periodic near horizon current model excludes these winding orbits, and obtain selection rules and closed form expressions for finite monodromy resolved orbital functionals. A unitary realization on the full infinite dimensional orbit is not constructed, and the interpretation of the cyclic sectors in quantum gravity remains conjectural.}


\begin{document}
\maketitle
\flushbottom

%======================================================================
\section{Introduction}\label{sec:intro}

The asymptotic symmetry algebra of three dimensional gravity without cosmological constant at null infinity is the centrally extended semidirect sum of the vector fields on the circle with the abelian ideal of supertranslations, with central charge $c_2=3/G$ in the mixed sector~\cite{Barnich:2006av}. Inducing data for representations of the corresponding BMS$_3$ group are organized by its coadjoint orbits. Induced representations for several sectors were studied by Barnich and Oblak~\cite{Barnich:2014kra,Barnich:2015uva}, their characters were computed in~\cite{Oblak:2015sea}, and a systematic account is given in~\cite{Oblak:2017ptc}. The vacuum character reproduces the one loop partition function of flat gravity~\cite{Barnich:2015mui}. Related modules and geometric actions were studied in~\cite{Campoleoni:2016vsh,Merbis:2019wgk}.

The supermomentum orbits are coadjoint orbits of the Virasoro group. Their classification goes back to Lazutkin and Pankratova~\cite{Lazutkin:1975}, Segal~\cite{Segal:1981ap} and Witten~\cite{Witten:1987ty}, and was organized in terms of the monodromy of a Hill operator by Balog, Feh\'er and Palla~\cite{Balog:1997zz}. The orbits of constant supermomentum describe the vacuum and the massive particles, and they have been the main focus of the representation theory~\cite{Barnich:2014kra,Oblak:2017ptc}. Maxfield classified the Schwarzian theories defined by integrals over arbitrary Virasoro orbits and computed their path integrals exactly~\cite{Maxfield:2026men}. For the hyperbolic orbits $T_{n,\Delta}$ he gave the general representative determined by a stabilizer vector field with $2n$ simple zeros, the solutions of the Hill equation and their monodromy, and he took the gauge group to be the one parameter subgroup generated by the stabilizer field. Mao and Mao constructed the one loop modules of the constant Virasoro and BMS$_3$ orbits~\cite{Mao:2026bms}, and Garbarz and Leston quantized BMS$_3$ orbits perturbatively around constant representatives~\cite{Garbarz:2015lua}. Ruzziconi and West found string like carriers in extended BMS$_3$ representations~\cite{Ruzziconi:2026ext}, and holonomy methods give access to one loop data in flat space~\cite{Bagchi:2023ilg,Pannier:2025wil}. In four dimensions, explicit BMS particles and representations with generic supermomentum were studied in~\cite{Bekaert:2024uuy,Bekaert:2025gen}.

A second line of work concerns the global group $B(2,1)$, in which the supertranslations form the Hilbert space of square integrable functions on the circle and the superrotations are restricted to the Lorentz group $\SL$~\cite{Melas:2017jzb,Melas:2021hil}. Its nonzero little groups are compact and, up to conjugacy, equal to $SO(2)$ or to a cyclic group $C_{2m}$ of even order. In the companion manuscript~\cite{Melas:2026iwe} the induced model for a sector with little group $C_{2m}$ and character $r\in\Z/2m\Z$ obeys an exact finite shift equation instead of a differential spin equation, and its Poincar\'e content is shown to be a ladder of helicities with multiplicity one.

We ask whether this finite cyclic structure is realized in the full superrotation group of gravity, and whether it carries information beyond a classical periodicity of the supermomentum. The answer is positive for the hyperbolic orbits with winding, and the reason is the structure of their stabilizer. In the spin double cover it is the direct product $\R\times C_{2m}$, so its characters carry a continuous label $s$ and an independent discrete label $r$. On the constant orbits the same cyclic group lies inside the circle of rotations, and its character is fixed by the spin. The label $r$ is therefore an independent character of the inducing stabilizer for these nonconstant orbits. Whether it is realized as a physical quantum gravitational superselection label requires a full quantization, which is not constructed here.

Our results are as follows. Using the developing map of the Hill operator we identify the stabilizer of any supermomentum whose lifted monodromy has nonzero translation number with the centralizer of that monodromy modulo the monodromy itself. This gives a self contained derivation of Witten's result $\R\times\Z_n$ for $T_{n,\Delta}$ and recovers the stabilizers of the exceptional constants. For every point of $T_{n,\Delta}$ or $T_{n,\pm}$ with $n\ge2$ the stabilizer meets the Lorentz group in a finite cyclic group of rigid rotations, up to conjugation. The points invariant under the rotation $R_m$ through $2\pi/m$ exist if and only if $m$ divides $n$, they form an infinite dimensional family, and on it the intersection is exactly $\langle R_m\rangle$. We give a direct proof that the parabolic orbits with winding have the same stabilizer structure, and we prove that the cyclic label is independent of the continuous data exactly for the hyperbolic and parabolic orbits with winding. Restriction of formal smooth equivariant sections to the Lorentz orbit through an invariant point intertwines the smooth actions and yields the finite cyclic equivariance condition for the shifted supermomentum $2\pi(q+\kappa/2)$. Separately, the corresponding finite cyclic $B(2,1)$ global branch is constructed as a unitary representation, without asserting that it is a closed subrepresentation of a unitary representation on the full orbit. When the projected momentum is timelike, the Poincar\'e content of the resulting global branch is the ladder $h=r/2+mn$, the Poincar\'e mass must include the vacuum shift $\kappa/2$, and on the family of invariant points the projected rest energy takes every real value. The Bondi data define locally flat winding BMS geometries admitting two commuting Killing vectors and with hyperbolic Lorentz holonomy of rapidity $\Delta$. Their global causal and regularity properties are not classified here. The smooth real periodic near horizon current model of~\cite{Afshar:2016kjj} does not produce these winding orbits. Monodromy resolved twisted amplitudes, defined as finite diagonal integrals of convolution kernels, reduce to elliptic orbital integrals and obey parity and divisibility selection rules. For test functions that depend on the norm of the matrix we evaluate them in closed form. For orientation odd test functions the dependence on $r$ is the universal factor $m-r$, and the amplitudes distinguish all $2m$ sectors. Finally, under the explicitly stated smooth principal bundle hypothesis, the fundamental group of the orbit is $\Z$. For inducing characters trivial on the connected stabilizer, their cyclic phase is the holonomy of the associated flat line bundle, giving a formal analogy with a $\theta$ angle.

\begin{figure}[t]
\centering
\begin{tikzpicture}[node distance=4mm and 5mm,box/.style={draw,rounded corners=2pt,align=center,font=\small,inner sep=4pt,minimum height=11mm},>={Stealth[length=2mm]}]
\node[box] (a) {orbit\\ $T_{n,\Delta}$, $T_{n,\pm}$};
\node[box,right=of a] (b) {stabilizer\\ $\R\times\Z_n$};
\node[box,right=of b] (c) {intersection\\ with $G_0$: $\langle R_m\rangle$};
\node[box,right=of c] (d) {cyclic label\\ $r\in\Z/2m\Z$};
\node[box,right=of d] (e) {Poincar\'e content\\ $h=\frac r2+mn$};
\draw[->] (a)--(b); \draw[->] (b)--(c); \draw[->] (c)--(d); \draw[->] (d)--(e);
\node[font=\scriptsize,below=1mm of b] {Thm.~\ref{thm:stabilizer}, Props.~\ref{prop:hyperbolic},~\ref{prop:parabolic}};
\node[font=\scriptsize,below=1mm of c] {Thm.~\ref{thm:intersection}};
\node[font=\scriptsize,below=1mm of d] {Thm.~\ref{thm:independence}};
\node[font=\scriptsize,below=1mm of e] {Thm.~\ref{thm:ladder}};
\end{tikzpicture}
\caption{Logical structure of the main results. The stabilizer of an orbit with winding $n$ is a centralizer of the lifted monodromy. At points with rotational symmetry of order $m$ it meets the Lorentz group in $\langle R_m\rangle$, whose lift $C_{2m}$ carries the cyclic label $r$, independent of the continuous label. When the projected momentum is timelike, restriction to the global orbit turns $r$ into the offset of a ladder of helicities.}
\label{fig:chain}
\end{figure}
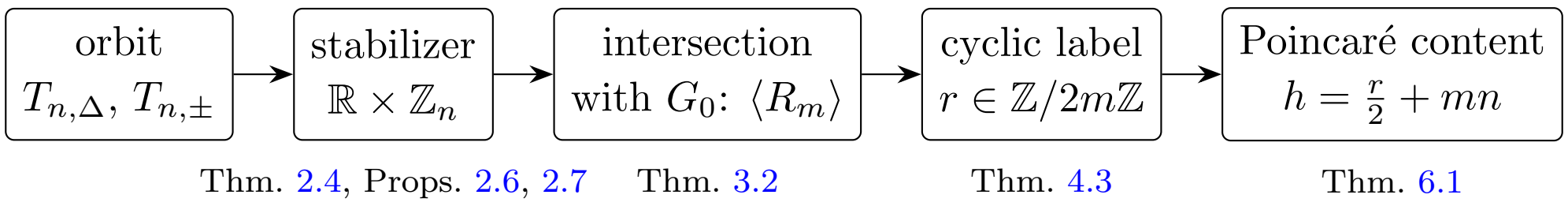

Figure~\ref{fig:chain} summarizes the logical chain from the orbit to the particle content.

The orbit classification by the lifted monodromy is classical~\cite{Lazutkin:1975,Segal:1981ap,Balog:1997zz}, the stabilizer structure is due to Witten~\cite{Witten:1987ty}, and the representative, Hill solutions and monodromy of $T_{n,\Delta}$ appear in~\cite{Witten:1987ty,Maxfield:2026men}. We therefore keep this material short. The closest works are~\cite{Witten:1987ty,Balog:1997zz,Maxfield:2026men}, and we compare our results with them claim by claim. Witten determines the stabilizers of all Virasoro orbits as subgroups of the diffeomorphism group, and Balog, Feh\'er and Palla classify the orbits through the conjugacy class of the lifted monodromy and relate them to the global Liouville equation. Neither work considers the BMS$_3$ extension, the intersection of the stabilizer with the Lorentz group, or the characters of its group of components. Maxfield computes path integrals over $T_{n,\Delta}$ with the identity component of the stabilizer as gauge group and does not use the discrete factor or its characters. Against this background, and to our knowledge, the following are new. The intersection with the Lorentz group and the classification of the rotationally invariant points (section~\ref{sec:cyclic}) use the discrete factor, which is not exploited in~\cite{Witten:1987ty,Balog:1997zz,Maxfield:2026men}. The independence of the cyclic label (section~\ref{sec:characters}) is a statement about the characters of that factor. The restriction to global orbits and the identification with the finite cyclic sectors of $B(2,1)$ (section~\ref{sec:restriction}) connect these data with the companion paper, while the BMS$_3$ representations constructed explicitly in~\cite{Barnich:2014kra,Oblak:2017ptc,Mao:2026bms,Garbarz:2015lua} are those of constant orbits or of expansions around them. The helicity ladder and the behaviour of the projected rest energy (section~\ref{sec:poincare}) concern the Poincar\'e content at invariant points, which is not discussed in these references. For the singular near horizon currents (section~\ref{sec:bulk}) the no go statement for smooth currents is a direct consequence of the Sturm separation theorem, and only the explicit singular description is new. The finite orbital amplitudes (section~\ref{sec:twisted}) are elementary orbital integrals on $\SL$ evaluated for the elements of $C_{2m}$, and the interpretation of the cyclic label as the holonomy of a flat line bundle (section~\ref{sec:topology}) holds under an explicit local triviality assumption. Finally,~\cite{Ruzziconi:2026ext,Bekaert:2024uuy,Bekaert:2025gen} concern extended or four dimensional settings. Additional relevant comparisons are the gauge theoretic analysis of Virasoro orbits, projective monodromy and winding by Blau and Youmans~\cite{Blau:2024}, and the explicit bosonization and orbit path integral constructions of Alekseev, Chekeres and Youmans~\cite{Alekseev:2024}. These works clarify the geometric and functional integral significance of hyperbolic and parabolic Virasoro orbits, but do not derive the finite cyclic global $B(2,1)$ branching and the finite orbital functionals considered here. Our contribution is not a new classification of the Virasoro stabilizers, nor the elementary character theory of a direct product by itself. Rather, it is the explicit connection of the discrete component character to the global Lorentz restriction, together with its selection rules and orbital integrals. This comparison is based on our reading of the cited works. We do not attempt a unitary quantization of the nonconstant orbit, which requires quasi invariant measures on an infinite dimensional homogeneous space. Our statements are formulated on smooth sections or on finite dimensional global orbits, and we do not claim that the restriction map has dense image. Accordingly, when we speak of BMS$_3$ representations and their quantum numbers we mean their inducing data and the smooth restrictions of equivariant sections to global orbits, and the helicity ladder refers to the timelike projected Poincar\'e content at invariant points.

This paper is organized as follows. Section~\ref{sec:hill} recalls the Hill operator picture and derives the stabilizers. Section~\ref{sec:cyclic} determines the intersections with the Lorentz group, the invariant points and the spin covers. Section~\ref{sec:characters} proves the independence of the cyclic label. Section~\ref{sec:restriction} constructs the restriction to the global orbits and derives the finite shift equation. Section~\ref{sec:poincare} gives the Poincar\'e content and the mass. Section~\ref{sec:bulk} describes the bulk geometries and proves the no go theorem for soft hair. Section~\ref{sec:twisted} treats twisted amplitudes, section~\ref{sec:topology} the topology of the orbit, and section~\ref{sec:discussion} open problems. Appendix~\ref{app:B21} explains how the finite cyclic little groups and the finite shift equation arise in $B(2,1)$, and appendix~\ref{app:tech} collects technical lemmas and the list of checks, which are reproduced by the supplementary reproducibility package.

%======================================================================
\section{Hill operators and stabilizers}\label{sec:hill}

This section collects the geometric input. We first recall the coadjoint action, the Hill operator and the developing map, and then derive the stabilizers and introduce the representative of $T_{m,\Delta}$ used throughout.

We fix the notation used in the whole paper. The superrotation group is $G=\Diff$, with universal cover $\widetilde G$ and connected $N$ fold covers $G^{(N)}=\widetilde G/\langle T^N\rangle$. The Lorentz group is $G_0=\PSL\subset G$, and its preimage in the spin double cover $G^{(2)}$ is $\SL$. The symbols $T_{n,\Delta}$ and $T_{n,\pm}$ denote the hyperbolic and parabolic orbit types of~\cite{Balog:1997zz}, with winding $n$, and $\Om_{m,\Delta}$ denotes the particular hyperbolic orbit of the representative $p_{m,a}$ introduced below, so that $\Om_{m,\Delta}=T_{m,\Delta}$. The integer $n$ is always a winding and $m$ the order of a rotational symmetry, with $m\mid n$. Sections~\ref{sec:hill} and~\ref{sec:cyclic} work in $G$ and $\widetilde G$. From section~\ref{sec:covers} on we work in $G^{(2)}$, where $\gamma=k(\pi/m)$ is the image of the lifted rotation $\widetilde R_m$, $\gamma_n$ generates the cyclic factor of a stabilizer on an orbit of winding $n$, the continuous label is $s\in\R$ and the cyclic label is $r\in\Z/2m\Z$, or $r_n\in\Z/2n\Z$ for winding $n$. On the universal cover the cyclic label becomes an angle $\alpha$, and $s'$ denotes a weight of the rotation group $K$.

\subsection{Coadjoint action and developing map}\label{sec:coadjoint}

Let $G=\Diff$, and realize its universal cover as $\widetilde G=\{F\in C^\infty(\R,\R):F'>0,\ F(\varphi+2\pi)=F(\varphi)+2\pi\}$. The translation $T(\varphi)=\varphi+2\pi$ is central and $G=\widetilde G/\langle T\rangle$. The set $\widetilde G$ is convex, hence contractible, and $\pi_1(G)\cong\Z$ is generated by the loop of rotations. A supermomentum $p\,d\varphi^2$ transforms as
\begin{equation}\label{eq:coadfinite}
g\cdot p=\big(p\circ g^{-1}\big)\big((g^{-1})'\big)^2-\kappa\,\Sch(g^{-1}),\qquad\Sch(h)=\frac{h'''}{h'}-\frac32\Big(\frac{h''}{h'}\Big)^2,
\end{equation}
with $\kappa=c_2/(24\pi)$, and infinitesimally $\delta_up=up'+2u'p-\kappa u'''$. The constants $p=-\kappa n^2/2$ are invariant under the $n$ fold cover $\PSL^{(n)}$, and $n=1$ gives the Minkowski vacuum $p_{\rm vac}=-\kappa/2$, whose stabilizer is the Lorentz group $G_0=\PSL$ acting by Möbius transformations in the angular coordinate. The geometry of an orbit is encoded in the Hill operator $L_p=\partial_\varphi^2-Q_p$ with $Q_p=p/(2\kappa)$.

The next lemma, proved by direct differentiation in appendix~\ref{app:tech}, expresses~\eqref{eq:coadfinite} in terms of solutions.

\begin{lemma}\label{lem:covariance}
Let $g\in G$ with lift $F$ and set $h=F^{-1}$. If $L_p\psi=0$, then $\widetilde\psi=(\psi\circ h)(h')^{-1/2}$ satisfies $L_{g\cdot p}\widetilde\psi=0$, and Wronskians $W(\psi_1,\psi_2)=\psi_1\psi_2'-\psi_1'\psi_2$ are preserved.
\end{lemma}

\begin{proof}
By~\eqref{eq:coadfinite}, $Q_{g\cdot p}=(Q_p\circ h)h'^2-\frac12\Sch(h)$, and appendix~\ref{app:tech} gives $\widetilde\psi''-Q_{g\cdot p}\widetilde\psi=\big((\psi''-Q_p\psi)\circ h\big)(h')^{3/2}$. In $\widetilde\psi_1\widetilde\psi_2'-\widetilde\psi_1'\widetilde\psi_2$ the terms containing $h''$ cancel, so $W(\widetilde\psi_1,\widetilde\psi_2)=W(\psi_1,\psi_2)\circ h$, which is constant. This completes the proof.
\end{proof}

A basis $\Psi=(\psi_1,\psi_2)$ of solutions with $W=1$ is called normalized. Since $\Psi$ never vanishes we can write $\Psi=\varrho(\cos\theta,\sin\theta)$ with $\varrho>0$ and a continuous angle $\theta$, the developing map of the projective structure defined by $p$~\cite{Ovsienko:2004}. The following lemma is its basic property.

\begin{lemma}\label{lem:developing}
For a normalized basis $\theta'=\varrho^{-2}>0$, $\psi_1=\cos\theta/\sqrt{\theta'}$, $\psi_2=\sin\theta/\sqrt{\theta'}$ and $Q_p=-\theta'^2-\frac12\Sch(\theta)$. Conversely, every $\theta$ with $\theta'>0$ defines in this way a normalized basis for $Q=-\theta'^2-\frac12\Sch(\theta)$.
\end{lemma}

\begin{proof}
One has $W=\varrho^2\theta'$, so normalization gives $\theta'=\varrho^{-2}$. The functions $\cos$ and $\sin$ form a normalized basis for $\psi''=-\psi$. The computation in the proof of Lemma~\ref{lem:covariance}, applied with $h=\theta$, shows that $\cos\theta/\sqrt{\theta'}$ and $\sin\theta/\sqrt{\theta'}$ form a normalized basis for $Q=-\theta'^2-\frac12\Sch(\theta)$. A potential is determined by any nonvanishing solution through $Q=\psi''/\psi$, and the two solutions have no common zero, so $Q_p$ has the stated form. This completes the proof.
\end{proof}

Since $Q_p$ is periodic, $\Psi(\varphi+2\pi)$ is again a normalized basis, and $\Psi(\varphi+2\pi)=M\Psi(\varphi)$ for a unique monodromy $M\in\SL$. The universal cover $\tSL$ acts faithfully on the line of angles, and there is a unique lift $\widetilde M$ with
\begin{equation}\label{eq:liftedmonodromy}
\theta(\varphi+2\pi)=\widetilde M\big(\theta(\varphi)\big).
\end{equation}
A change of normalized basis conjugates $\widetilde M$, and by Lemma~\ref{lem:covariance} its conjugacy class is constant on the orbit. The centre of $\tSL$ is generated by $\zeta(x)=x+\pi$. Every $C\in\tSL$ has a translation number $\tau(C)=\lim_n(C^n(x)-x)/n$, which is a conjugation invariant with $\tau(\zeta^kC)=k\pi+\tau(C)$, and $\tau(C)=0$ if $C$ has a fixed point (appendix~\ref{app:tech}). Since the solution $\cos(\theta-\theta_0)/\sqrt{\theta'}$ vanishes exactly when $\theta\in\theta_0+\frac\pi2+\pi\Z$, every solution has asymptotically $|\tau(\widetilde M)|/\pi$ zeros per period. Since $\theta'>0$, $\widetilde M(x)>x$ for $x=\theta(\varphi)$, so $\tau(\widetilde M)\ge0$ for every $p$. We call this number the winding of the orbit. For the hyperbolic orbits it is the integer $n$ in $T_{n,\Delta}$. We call an orbit topologically trivial if $|\tau(\widetilde M)|<\pi$, which holds for the orbits of the constants $p_0>-\kappa/2$, among them the conical defects and the flat space cosmologies. By~\cite{Witten:1987ty} and~\cite{Balog:1997zz}, sections~4 and~5, the energy $\int p\,d\varphi$ is bounded from below on the orbits of the constants $p_0\ge-\kappa/2$, and it is unbounded from below on every $T_{n,\Delta}$ with $n\ge1$, as we also show directly in Lemma~\ref{lem:unbounded}.

The Poincar\'e translations are the supertranslations $f_0=1$, $f_1=\cos\varphi$, $f_2=\sin\varphi$, and $\sltwo$ is spanned by the vector fields with the same coefficients. The next lemma shows that the Poincar\'e momentum must be computed with the vacuum subtracted.

\begin{lemma}\label{lem:projection}
For $u\in\sltwo$ one has $\delta_up=u\bar p'+2u'\bar p$ with $\bar p=p+\kappa/2$. Hence $P_\mu(p)=\int_0^{2\pi}(p+\frac\kappa2)f_\mu\,d\varphi$ intertwines the action of $G_0$ with the coadjoint action on $\sltwo^*\cong\R^{2,1}$, and $P_0^2-P_1^2-P_2^2$ is invariant.
\end{lemma}

\begin{proof}
For $u\in\sltwo$, $u'''=-u'$, so $-\kappa u'''=2u'\frac\kappa2$. For $f$ in the same span, integration by parts gives $\delta_u\int\bar pf=\int\bar p\,(u'f-uf')$, and $uf'-u'f$ is the Lie bracket of two elements of $\sltwo$. This proves equivariance under the Lie algebra, and under $G_0$ by connectedness. The Killing form in the basis $(1,\cos\varphi,\sin\varphi)$ is $\mathrm{diag}(-2,2,2)$, so the dual form $P_0^2-P_1^2-P_2^2$ is invariant up to sign. This completes the proof.
\end{proof}

Physically, $P_0$ is the energy measured from the Minkowski vacuum. A constant $p_0>-\kappa/2$ has $P=(2\pi(p_0+\kappa/2),0,0)$, the rest momentum of the massive particles of~\cite{Barnich:2014kra}.

\subsection{Stabilizers and a hyperbolic representative}\label{sec:stabilizers}

Let $H_p\subset G$ be the stabilizer of $p$ and $\widetilde H_p\subset\widetilde G$ its preimage. The developing map turns $\widetilde H_p$ into a centralizer, as the following theorem shows.

\begin{theorem}\label{thm:stabilizer}
If $\tau(\widetilde M)\neq0$, then $\theta$ is a diffeomorphism of $\R$, and $F\mapsto C_F=\theta\circ F\circ\theta^{-1}$ is an isomorphism from $\widetilde H_p$ onto the centralizer $Z(\widetilde M)$ of $\widetilde M$ in $\tSL$ sending $T$ to $\widetilde M$. Hence $H_p\cong Z(\widetilde M)/\langle\widetilde M\rangle$.
\end{theorem}

\begin{proof}
Since $\theta$ is increasing, $\widetilde M\theta(\varphi)=\theta(\varphi+2\pi)>\theta(\varphi)$, and since $\widetilde M$ has no fixed point, the continuous function $\widetilde M(x)-x$ is positive for all $x$. If $\theta(\R)$ were bounded above, the increasing sequence $\theta(\varphi+2\pi k)=\widetilde M^k\theta(\varphi)$ would converge to a fixed point of $\widetilde M$. Hence $\theta(\R)$ is unbounded above, and similarly below, so $\theta$ is a diffeomorphism of $\R$.

Let $F\in\widetilde H_p$ and $h=F^{-1}$. By Lemma~\ref{lem:covariance}, $(\Psi\circ h)(h')^{-1/2}$ is a normalized basis for $L_p$, hence equal to $B\Psi$ for a unique $B\in\SL$. Its angle is $\theta\circ h$, and it is also a continuous lift of the angle of $B\Psi$, so $\theta\circ h=\widetilde B\circ\theta$ for a lift $\widetilde B$. Thus $C_F=\widetilde B^{-1}\in\tSL$. The map $F\mapsto C_F$ is an injective homomorphism, $C_T=\widetilde M$ by~\eqref{eq:liftedmonodromy}, and $C_F$ commutes with $\widetilde M$ because $F$ commutes with $T$. Conversely, let $C\in Z(\widetilde M)$ and $F=\theta^{-1}\circ C\circ\theta$. Then $F(\varphi+2\pi)=\theta^{-1}(C\widetilde M\theta(\varphi))=\theta^{-1}(\widetilde M\theta(F(\varphi)))=F(\varphi)+2\pi$, so $F\in\widetilde G$. Let $g$ be its image. The transported basis for $L_{g\cdot p}$ has angle $C^{-1}\circ\theta$. Every $\phi\in\tSL$ satisfies $\Sch(\phi)=2(1-\phi'^2)$ (appendix~\ref{app:tech}), and with $\phi=C^{-1}$ the cocycle identity gives
\[
Q_{g\cdot p}=-\theta'^2\Big(\phi'^2+\tfrac12\Sch(\phi)\Big)\circ\theta-\tfrac12\Sch(\theta)=-\theta'^2-\tfrac12\Sch(\theta)=Q_p .
\]
Hence $F\in\widetilde H_p$ with $C_F=C$. Dividing by $\langle T\rangle$ gives the last statement. This completes the proof.
\end{proof}

The isomorphism is also a homeomorphism, when $\widetilde H_p$ carries the topology induced by $\widetilde G$ and $Z(\widetilde M)$ its Lie group topology. Indeed, the Lie group topology of $\tSL$ coincides with the compact open topology of its action on the line of angles, and $C_F=\theta\circ F\circ\theta^{-1}$ depends continuously on $F$ for the topology of uniform convergence on compact sets, which is weaker than the topology of $\widetilde G$. Conversely, $F=\theta^{-1}\circ C_F\circ\theta$ and all its derivatives depend continuously on $C_F$, because $C_F$ ranges over a finite dimensional Lie group acting smoothly on $\R$. Statements about identity components and Lie group structures of stabilizers refer to this topology.

Theorem~\ref{thm:stabilizer} also covers the central case. For the exceptional constant $p=-\kappa n^2/2$, the normalized basis $\sqrt{2/n}\,(\cos\frac{n\varphi}2,\sin\frac{n\varphi}2)$ has $\theta=n\varphi/2$, so $\widetilde M=\zeta^n$ and the theorem gives $H_p\cong\tSL/\langle\zeta^n\rangle=\PSL^{(n)}$, as it should. The same argument gives a short proof of the classical fact that the lifted monodromy classifies the orbits~\cite{Lazutkin:1975,Segal:1981ap,Balog:1997zz}, in the form we need.

\begin{corollary}\label{cor:classification}
If $\tau(\widetilde M_p),\tau(\widetilde M_{p'})\neq0$, then $p'\in G\cdot p$ if and only if $\widetilde M_{p'}$ is conjugate to $\widetilde M_p$.
\end{corollary}

\begin{proof}
One direction was shown above. Conversely, after a change of basis $\widetilde M_{p'}=\widetilde M_p=\widetilde M$. Let $\theta_p$ and $\theta_{p'}$ be the corresponding angles, which are diffeomorphisms of $\R$ by Theorem~\ref{thm:stabilizer}. The map $F=\theta_{p'}^{-1}\circ\theta_p$ satisfies
\[
F(\varphi+2\pi)=\theta_{p'}^{-1}\big(\widetilde M\theta_p(\varphi)\big)=\theta_{p'}^{-1}\big(\widetilde M\theta_{p'}(F(\varphi))\big)=F(\varphi)+2\pi ,
\]
so $F\in\widetilde G$. By Lemma~\ref{lem:covariance} its image $g\in G$ transports the basis of $L_p$ to a normalized basis of $L_{g\cdot p}$ with angle $\theta_p\circ F^{-1}=\theta_{p'}$, and Lemma~\ref{lem:developing} gives $Q_{g\cdot p}=-\theta_{p'}'^2-\frac12\Sch(\theta_{p'})=Q_{p'}$. This completes the proof.
\end{proof}

For hyperbolic monodromy the centralizer is elementary. Let $A=\{\exp(tX)\}\subset\PSL$ be a hyperbolic one parameter group and $\widetilde A$ its lift through the identity.

\begin{proposition}\label{prop:hyperbolic}
If $|\tr M|>2$ and $\tau(\widetilde M)=k\pi$ with $k\ge1$, then after conjugation $\widetilde M=\exp(\lambda X)\zeta^k$ with $\lambda\neq0$, $Z(\widetilde M)=\widetilde A\times\langle\zeta\rangle$, and $H_p\cong\R\times\Z_{k}$. Its identity component is the flow of a stabilizer field with $2k$ simple zeros, and $\{0\}\times\Z_{k}$ is the set of its elements of finite order.
\end{proposition}

\begin{proof}
The image $\bar M$ of $M$ in $\PSL$ is hyperbolic, so after conjugation $\bar M\in A$, whose centralizer in $\PSL$ is $A$. The preimage of $A$ in $\tSL$ is $\widetilde A\times\langle\zeta\rangle$, a direct product because elements of $\widetilde A$ have fixed points and powers $\zeta^j\neq1$ do not. Writing $\widetilde M=\exp(\lambda X)\zeta^j$ we get $\tau(\widetilde M)=j\pi$, so $j=k$, and $\lambda\neq0$ since $\bar M\neq1$. An element commuting with $\widetilde M$ has image commuting with $\bar M$, hence lies in $\widetilde A\times\langle\zeta\rangle$, which is abelian. The homomorphism $(t,j)\mapsto(t-j\lambda/k,\ j\bmod k)$ from $\R\times\Z$ onto $\R\times\Z_{k}$ has kernel $\langle(\lambda,k)\rangle$, which gives $H_p$ by Theorem~\ref{thm:stabilizer}. The identity component consists of the maps $\theta^{-1}\circ\exp(tX)\circ\theta$, which form the flow of the vector field $v=(\hat X\circ\theta)/\theta'$, where $\hat X$ is the vector field on the line of angles generating $\exp(tX)$. The line of angles parametrizes the rays of $\R^2$, and $\zeta$ is the rotation by $\pi$, which maps a ray to its opposite. The field $\hat X$ vanishes exactly on the set $Z$ of angles of the rays in the two eigenlines of $A$. Each eigenline contains two opposite rays, so $Z$ is the union of two arithmetic progressions of step $\pi$. The zeros are simple because the eigenvalues of $A$ differ from one, so the derivative of $\exp(tX)$ at a fixed ray differs from one for $t\neq0$. The zeros of $v$ in $[\varphi,\varphi+2\pi)$ correspond under the increasing map $\theta$ to the points of $Z$ in $[y,\widetilde My)$ with $y=\theta(\varphi)$. The set $Z$ is invariant under $\exp(\lambda X)$, which fixes each of its points, and under $\zeta$, so $\widetilde M$ acts on $Z$ as the translation by $k\pi$. If $z_0$ is the smallest point of $Z$ with $z_0\ge y$, then $z_0+k\pi=\widetilde Mz_0$ is the smallest point of $Z$ with $z\ge\widetilde My$, because $\widetilde M$ is increasing and maps $Z$ onto itself. Hence $Z\cap[y,\widetilde My)=Z\cap[z_0,z_0+k\pi)$, which contains exactly $2k$ points. Thus $v$ has exactly $2k$ zeros per period, all simple since $\theta'>0$. The elements of finite order of $\R\times\Z_{k}$ are those of $\{0\}\times\Z_{k}$. This completes the proof.
\end{proof}

The same argument applies to parabolic monodromy, where $\widetilde A$ is replaced by the lift $\widetilde N$ of a parabolic one parameter group $N=\{\exp(tY)\}$. This gives a direct proof of the classical description of the stabilizers of the parabolic orbits $T_{n,\pm}$~\cite{Witten:1987ty,Balog:1997zz}, together with the structure of the zeros needed below.

\begin{proposition}\label{prop:parabolic}
If $|\tr M|=2$, $M\neq\pm I$ and $\tau(\widetilde M)=k\pi$ with $k\ge1$, then after conjugation $\widetilde M=\exp(\pm Y)\zeta^k$, $Z(\widetilde M)=\widetilde N\times\langle\zeta\rangle$, and $H_p\cong\R\times\Z_{k}$. Its identity component is the flow of a stabilizer field with $k$ double zeros per period, and $\{0\}\times\Z_{k}$ is the set of its elements of finite order.
\end{proposition}

\begin{proof}
The image $\bar M$ is a nontrivial parabolic element, and after conjugation in $\PSL$ it equals $\exp(\pm Y)$. A matrix commuting with a nontrivial unipotent matrix is $\pm$ a unipotent matrix with the same fixed line, so the centralizer of $\bar M$ in $\PSL$ is $N$. Elements of $\widetilde N$ have fixed points, so the preimage of $N$ is the direct product $\widetilde N\times\langle\zeta\rangle$, and the argument of Proposition~\ref{prop:hyperbolic} gives $Z(\widetilde M)$ and $H_p\cong\R\times\Z_{k}$, with $\pm1$ in place of $\lambda$. The generator $\hat Y$ of $\widetilde N$ on the line of angles vanishes on the angles of the two opposite rays of the single line fixed by $N$, which form one arithmetic progression of step $\pi$. For $Y=E_{12}$ the angle of $(\cos x+t\sin x,\sin x)$ has derivative $-\sin^2x$ at $t=0$, so $\hat Y(x)=-\sin^2x$ vanishes exactly to second order. The translation number is $\tau(C)=\lim_n(C^n(x)-x)/n$ as in section~\ref{sec:coadjoint}. It is invariant under conjugation in $\tSL$, so neither $k$ nor the quotient $Z(\widetilde M)/\langle\widetilde M\rangle$ depends on the conjugating lift, which only replaces $\widetilde N$ by a conjugate subgroup. The counting argument of Proposition~\ref{prop:hyperbolic} gives $k$ zeros of $v=(\hat Y\circ\theta)/\theta'$ per period, each a double zero. This completes the proof.
\end{proof}

We now fix $m\ge1$, $\eta>0$, $a=-\tanh\eta$, $C=\cosh\eta$ and $\Delta=2\pi m\sinh\eta$, and set
\begin{equation}\label{eq:rep}
u_{m,a}=\sin(m\varphi)\big(1+a\sin(m\varphi)\big),\qquad
p_{m,a}=-\frac{\kappa m^2}{2}\,\frac{1+6a\sin(m\varphi)+4a^2\sin^2(m\varphi)}{\big(1+a\sin(m\varphi)\big)^2}.
\end{equation}
The field $u_{m,a}$ has $2m$ simple zeros $j\pi/m$ with $|u_{m,a}'|=m$ there, both functions are invariant under $R_m(\varphi)=\varphi+2\pi/m$, and $p_{m,a}(\varphi)=m^2p_{1,a}(m\varphi)$. A stabilizer field $u$ with simple zeros and $|u'|=\lambda$ at the zeros determines the smooth supermomentum $p=\kappa(2uu''-u'^2+\lambda^2)/(2u^2)$, with Hill solutions $\sqrt{|u|}\exp(\pm\frac\lambda2\int u^{-1})$~\cite{Maxfield:2026men}. One checks that~\eqref{eq:rep} is the case $u=u_{m,a}$, $\lambda=m$ of this formula, and here the integral of $u^{-1}$ can be done explicitly. The next proposition records the result.

\begin{proposition}\label{prop:floquet}
With $x=m\varphi$ and $F(x)=\int_0^x\frac{dt}{1+a\sin t}$, the functions
\[
\psi_1=\sqrt2\sin\tfrac x2\sqrt{1+a\sin x}\,e^{-aF/2},\qquad\psi_2=\sqrt2\cos\tfrac x2\sqrt{1+a\sin x}\,e^{aF/2}
\]
solve $L_{p_{m,a}}\psi=0$, with $\psi_1\psi_2=u_{m,a}$, $W(\psi_2,\psi_1)=\psi_2\psi_1'-\psi_2'\psi_1=m$ in the convention of Lemma~\ref{lem:covariance}, and
\begin{equation}\label{eq:mono}
\psi_{1,2}(\varphi+2\pi)=(-1)^me^{\pm\Delta/2}\psi_{1,2}(\varphi),\qquad \tr M=(-1)^m2\cosh\frac\Delta2,\qquad |\tau(\widetilde M)|=m\pi .
\end{equation}
\end{proposition}

\begin{proof}
Since $\frac{d}{dx}\big(\log|\tan\frac x2|-aF\big)=\frac1{\sin x(1+a\sin x)}$, the functions have the form $\sqrt{|u|}e^{\pm\chi/2}$ with $\chi'=m/u$ in $\varphi$, and with $u=u_{m,a}$ one finds $\psi''/\psi=(\log|\psi|)''+(\log|\psi|)'^2=\frac{u''}{2u}-\frac{u'^2}{4u^2}+\frac{m^2}{4u^2}$, because the terms linear in $\chi'$ cancel by $\chi''=-u'\chi'/u$. By the formula above this equals $p_{m,a}/(2\kappa)$, and the identity extends across the zeros of $u$ by smoothness. The Wronskian is $W(\psi_2,\psi_1)=u\chi'=m$, because $\psi_1$ carries the factor $e^{+\chi/2}$ and $\psi_2$ the factor $e^{-\chi/2}$. Under $\varphi\mapsto\varphi+2\pi$, $\sin\frac x2$ and $\cos\frac x2$ change by $(-1)^m$ and $F$ by $2\pi mC$, and $-aC=\sinh\eta$. Each $\psi_j$ has $m$ zeros per period. This completes the proof.
\end{proof}

Since $(\psi_2,\psi_1)/\sqrt m$ is normalized, Proposition~\ref{prop:hyperbolic} applies with $k=m$. The stabilizer of $p_{m,a}$ is $H\cong\R\times\Z_m$, generated by the flow $\Phi_t$ of $u_{m,a}$ and by $R_m$, which has order $m$ and therefore generates the torsion subgroup. The orbit of $p_{m,a}$ is $T_{m,\Delta}$~\cite{Witten:1987ty,Balog:1997zz}, which we denote $\Om_{m,\Delta}$. Allowing $\eta<0$, which changes the sign of $a$ and leaves $|\Delta|$ unchanged, amounts to the rotation $\varphi\mapsto\varphi+\pi/m$, and as $\eta\to0$ the family degenerates onto the exceptional constant $-\kappa m^2/2$. There is no ambiguity in this limit: $p_{m,a}$ is a smooth function of $(a,\varphi)$ for $|a|<1$, so $p_{m,a}\to-\kappa m^2/2$ uniformly with all derivatives, while the orbits $\Om_{m,\Delta}$ themselves change type at $a=0$, since $\Delta\to0$ and $\widetilde M\to\zeta^m$. No $T_{n,\Delta}$ contains a constant, since constants have stabilizers $SO(2)$ or $\PSL^{(n)}$. Physically, these orbits describe configurations whose Lorentz holonomy is a boost, as for a flat space cosmology, but whose developing map winds $m$ times. Besides the hyperbolic flow, their stabilizer contains a group of rigid rotations, which is the origin of the quantum number studied below.

%======================================================================
\section{Cyclic points and the Lorentz group}\label{sec:cyclic}

The finite cyclic sectors of $B(2,1)$ are defined by little groups inside the Lorentz group. We determine how the stabilizers meet $G_0$, which points carry a nontrivial intersection, and how the intersection lifts to the spin covers.

\subsection{Intersections with the Lorentz group}\label{sec:intersections}

The rigid rotations form the maximal compact subgroup $K\subset G_0$, the stabilizer of the centre $o$ of the hyperbolic disc $\Hb^2$, and $R_m$ denotes the rotation through $2\pi/m$. We first show that compact intersections containing $R_m$ consist of rigid rotations.

\begin{lemma}\label{lem:compact}
Let $R_m\in H_q$ with $m\ge2$, and let $G_0\cap H_q$ be compact. Then $G_0\cap H_q$ consists of rigid rotations. It is $K$ if $q$ is constant and otherwise $\langle R_{m'}\rangle$, where $2\pi/m'$ is the smallest angle of a rotation preserving $q$, so that $m\mid m'$.
\end{lemma}

\begin{proof}
Averaging the Euclidean inner product over the compact preimage of $G_0\cap H_q$ in $\SL$ gives an invariant inner product, so $G_0\cap H_q\subset gKg^{-1}$, the stabilizer of $g\cdot o$. The nontrivial elliptic element $R_m$ fixes only $o$ in $\Hb^2$, hence $g\cdot o=o$ and $G_0\cap H_q\subset K$. A closed subgroup of $K$ is $K$, which happens exactly when $q$ is constant, or finite cyclic. This completes the proof.
\end{proof}

For the hyperbolic and parabolic orbits with winding, compactness is automatic, as the following theorem shows.

\begin{theorem}\label{thm:intersection}
Let $n\ge2$ and let $q$ lie in $\Om_{n,\Delta}$ or in a parabolic orbit $T_{n,\pm}$. Then $G_0\cap H_q$ is finite cyclic of order dividing $n$ and is conjugate in $G_0$ to a group of rigid rotations. If $q$ is invariant under $R_m$ with $m\ge2$, then $m\mid n$ and $G_0\cap H_q=\langle R_{m'}\rangle$ with $m\mid m'\mid n$.
\end{theorem}

\begin{proof}
By Propositions~\ref{prop:hyperbolic} and~\ref{prop:parabolic}, $H_q\cong\R\times\Z_n$, where $\R$ is the flow $\Phi^q_t$ of a field with $2n$ simple zeros, respectively $n$ double zeros. For $t\neq0$, $\Phi^q_t$ fixes these zeros and moves every other point, since the integral curves between consecutive zeros are nonconstant. A Möbius transformation different from the identity has at most two fixed points, and if it has two, it is hyperbolic and its derivative at them differs from one. In the hyperbolic case $\Phi^q_t$ has $2n\ge4$ fixed points. In the parabolic case it has $n\ge2$ fixed points, at which its derivative is one because the zeros are double. Hence $\Phi^q_t\notin G_0$ for $t\neq0$. If $h=\Phi^q_t\delta\in G_0\cap H_q$ with $\delta$ torsion, then $h^n=\Phi^q_{nt}\in G_0$, so $t=0$. Thus $G_0\cap H_q$ lies in the torsion subgroup $\Z_n$, and being finite it is conjugate into $K$ by the averaging argument above. If $q$ is invariant under $R_m$, then $R_m$ has order $m$ in $\Z_n$, so $m\mid n$, and Lemma~\ref{lem:compact} gives $\langle R_{m'}\rangle\subset\Z_n$ with $m\mid m'$. This completes the proof.
\end{proof}

For $q=p_{m,a}$ this gives $G_0\cap H=\langle R_m\rangle$, because $m\mid m'\mid m$. The theorem says more than this. The one parameter group of hyperbolic superrotations that preserves $q$ never meets the Lorentz group, so it cannot be mistaken for a Lorentz symmetry, while the discrete part of the stabilizer is seen by the Lorentz group exactly at the points where it acts by rigid rotations. For $n=1$ the stabilizer field has only two zeros and the argument fails, consistently with the fact that the torsion subgroup $\Z_1$ is trivial. In the parabolic case the counting of fixed points alone would not suffice for $n=2$, since a hyperbolic Möbius transformation also has two fixed points, and the argument uses in addition that a flow with double zeros is tangent to the identity there. The theorem therefore treats the two orbit types with winding on the same footing. Its conclusion fails for the orbits of the constants, whose stabilizers contain the whole circle $K$ or the group $\PSL^{(n)}$, so that the intersection with $G_0$ is not finite.

Physically, a nontrivial intersection means that the gravitational configuration $q$ has a finite rotational symmetry on the celestial circle, and that this symmetry is part of its little group. A quantum state built on $q$ may then transform under the rotation by a phase, and this phase is the quantum datum that we want to isolate. The set of points with a nontrivial intersection is invariant under $G_0$, so it is natural to describe it modulo the Lorentz group, which we do next.

\subsection{Invariant points and spin covers}\label{sec:family}

The invariant points are plentiful, and their stabilizers lift to the spin covers in a simple way. Define $\Lambda_m(w)(\varphi)=m^2w(m\varphi)$, so that $\Lambda_m(p_{1,a})=p_{m,a}$, and for $g\in G$ with lift $\widetilde g$ let $\hat g$ be the image of $\varphi\mapsto\widetilde g(m\varphi)/m$, which commutes with $R_m$. The classification of invariant points rests on a uniqueness property of roots in the universal cover, which we state first.

\begin{lemma}\label{lem:root}
Let $\lambda\neq0$ and $m\ge1$. If $\widetilde N\in\tSL$ satisfies $\widetilde N^m=\exp(\lambda X)\zeta^m$, then $\widetilde N=\exp(\lambda X/m)\zeta$.
\end{lemma}

\begin{proof}
The element $\exp(\lambda X)\zeta^m$ is hyperbolic with translation number $m\pi\neq0$, so by Proposition~\ref{prop:hyperbolic} its centralizer is $\widetilde A\times\langle\zeta\rangle$. Since $\widetilde N$ commutes with its own $m$th power, $\widetilde N=\exp(\mu X)\zeta^j$ for unique $\mu\in\R$ and $j\in\Z$. The group $\widetilde A\times\langle\zeta\rangle$ is isomorphic to $\R\times\Z$ through $(\mu,j)$, because the product is direct, so the central index $j$ is an integer determined by the element and not only a class modulo $m$. The map $x\mapsto x^m$ is $(\mu,j)\mapsto(m\mu,mj)$, which is injective, and $(m\mu,mj)=(\lambda,m)$ gives $\mu=\lambda/m$ and $j=1$. This completes the proof.
\end{proof}

The following proposition describes all invariant points.

\begin{proposition}\label{prop:family}
\begin{enumerate}[label=(\roman*),nosep]
\item $\Lambda_m(g\cdot w)=\hat g\cdot\Lambda_m(w)$ for all $g\in G$ and all supermomenta $w$.
\item The points of $\Om_{m,\Delta}$ invariant under $R_m$ are exactly the points of $\Lambda_m(\Om_{1,\Delta/m})$, where $\Om_{1,\Delta/m}$ is the orbit of $p_{1,a}$, whose rapidity is $2\pi\sinh\eta=\Delta/m$.
\item $\Om_{n,\Delta}$ contains points invariant under $R_m$ if and only if $m\mid n$.
\item For $m\ge2$ every $q\in\Lambda_m(\Om_{1,\Delta/m})$ satisfies $G_0\cap H_q=\langle R_m\rangle$.
\end{enumerate}
\end{proposition}

\begin{proof}
(i) The lift of $\hat g^{-1}$ is $\varphi\mapsto\widetilde h(m\varphi)/m$ with $\widetilde h=\widetilde g^{-1}$. Its derivative is $\widetilde h'(m\varphi)$ and its Schwarzian is $m^2\Sch(\widetilde h)(m\varphi)$, and inserting these into~\eqref{eq:coadfinite} gives the claim.
(ii) Let $\Psi$ be a normalized basis for $L_w$ in the variable $x$, with angle $\theta$ and lifted monodromy $\widetilde N$. Since the Hill potential of $\Lambda_m(w)$ is $m^2w(m\varphi)/(2\kappa)$, the pair $m^{-1/2}\Psi(m\varphi)$ solves the Hill equation of $\Lambda_m(w)$, and its Wronskian in $\varphi$ is $m^{-1}\cdot m=1$. Its angle is $\theta(m\varphi)$, and $\theta(m\varphi+2\pi m)=\widetilde N^m\theta(m\varphi)$, so the lifted monodromy of $\Lambda_m(w)$ is $\widetilde N^m$. By (i), $\Lambda_m$ maps the orbit of $p_{1,a}$ into the orbit of $\Lambda_m(p_{1,a})=p_{m,a}$, so $\Lambda_m(\Om_{1,\Delta/m})\subset\Om_{m,\Delta}$, and all its points are invariant under $R_m$. Conversely, let $q\in\Om_{m,\Delta}$ be invariant under $R_m$. Then $w(x)=m^{-2}q(x/m)$ is $2\pi$ periodic and $q=\Lambda_m(w)$. Choose a normalized basis for $L_{p_{1,a}}$, and use its rescaling, which is a normalized basis for $L_{p_{m,a}}$, to define $\widetilde M_{p_{m,a}}$, so that $\widetilde N_{p_{1,a}}^m=\widetilde M_{p_{m,a}}$. After conjugating this basis we may assume $\widetilde M_{p_{m,a}}=\exp(\lambda X)\zeta^{m}$, by Proposition~\ref{prop:hyperbolic}. By Corollary~\ref{cor:classification}, $\widetilde N_w^m=\widetilde M_q$ is conjugate to $\widetilde M_{p_{m,a}}$, and after a change of the basis for $L_w$ we may assume $\widetilde N_w^m=\widetilde M_{p_{m,a}}$. Lemma~\ref{lem:root}, applied to $\widetilde N_w$ and to $\widetilde N_{p_{1,a}}$, gives $\widetilde N_w=\widetilde N_{p_{1,a}}=\exp(\lambda X/m)\zeta$, whose translation number is $\pi\neq0$, and $w\in\Om_{1,\Delta/m}$ by Corollary~\ref{cor:classification}.
(iii) An element of order $m$ in $\R\times\Z_n$ lies in $\Z_n$, so $m\mid n$. Conversely $p_{n,a}$ is invariant under $R_m=R_n^{n/m}$.
(iv) follows from (ii) and Theorem~\ref{thm:intersection} with $n=m$. This completes the proof.
\end{proof}

The family $\Lambda_m(\Om_{1,\Delta/m})$ is the image of an infinite dimensional orbit under an injective linear map, so it is not a finite union of Lorentz orbits. Every point of it carries the same intersection with $G_0$ and, as shown in section~\ref{sec:restriction}, the same finite shift equation. Quantities that depend on the point, such as the Poincar\'e mass, must therefore be studied on the whole family. Part (i) has a simple meaning: the superrotations commuting with $R_m$ act on symmetric configurations as superrotations of one sector of angular size $2\pi/m$.

\label{sec:covers}The discrete label depends on the cover of the superrotation group on which the representation is defined. Let $G^{(N)}=\widetilde G/\langle T^N\rangle$ be the connected $N$ fold cover. The preimage of $G_0$ in $G^{(2)}$ is $\SL$, because the loop of rotations generates both $\pi_1(G_0)$ and $\pi_1(G)$. We write $k(\omega)\in SO(2)\subset\SL$ for the rotation matrix of angle $\omega$, which acts on the circle by $\varphi\mapsto\varphi+2\omega$, so that $k(\pi)=-I$ is the image of $T$ and $k(\pi/m)$ lifts $R_m$.

Let $\widetilde R_m(\varphi)=\varphi+2\pi/m$, so that $\widetilde R_m^m=T$. The preimage of the stabilizer of $p_{m,a}$ in $\widetilde G$ is generated by the lifted flow $\widetilde\Phi_t$ and by $\widetilde R_m$. They commute, because $u_{m,a}$ is invariant under $R_m$, and $\{\widetilde\Phi_t\}\cap\langle\widetilde R_m\rangle$ is trivial, because $\widetilde\Phi_t$ fixes the lifts of the zeros while $\widetilde R_m^j$ with $j\neq0$ has no fixed point. Hence
\begin{equation}\label{eq:Htilde}
\widetilde H=\{\widetilde\Phi_t\}\times\langle\widetilde R_m\rangle\cong\R\times\Z ,
\end{equation}
and dividing by $\langle T^N\rangle=\langle\widetilde R_m^{mN}\rangle$ proves the following statement.

\begin{proposition}\label{prop:covers}
Let $q\in\Om_{m,\Delta}$ be invariant under $R_m$, $m\ge2$. In $G^{(N)}$ its stabilizer is $H^{(N)}_q\cong\R\times C_{mN}$, with cyclic factor generated by the image $\gamma_N$ of $\widetilde R_m$. In the double cover
\begin{equation}\label{eq:C2m}
\SL\cap H^{(2)}_q=C_{2m}=\langle\gamma\rangle,\qquad\gamma=k(\pi/m),\qquad\gamma^m=-I .
\end{equation}
\end{proposition}

\begin{proof}
For $q=p_{m,a}$ the first statement is~\eqref{eq:Htilde}. In general $q=\hat g\cdot p_{m,a}$ by Proposition~\ref{prop:family}, and the lift $\widetilde g(m\varphi)/m$ of $\hat g$ commutes with $\widetilde R_m$, so conjugation by it maps the preimages of the stabilizers onto each other and fixes $\widetilde R_m$. The intersection $\SL\cap H^{(2)}_q$ is the preimage of $G_0\cap H_q=\langle R_m\rangle$, which is generated by $k(\pi/m)$, an element of order $2m$ with $k(\pi/m)^m=-I$. This completes the proof.
\end{proof}

On $G$ itself only bosonic states occur and the cyclic label takes $m$ values. On the double cover it takes $2m$ values, with parity fixed by the statistics, and on the universal cover it becomes a continuous phase. The next section makes this precise.

%======================================================================
\section{Discrete characters}\label{sec:characters}

The stabilizers found above are the little groups of the induced representations. We determine their characters, show that the cyclic label is not affected by the central extension, and prove that it is independent of the continuous label exactly for the hyperbolic and parabolic orbits with winding.

\subsection{Characters and central extension}\label{sec:chars}

In the induced representations the supertranslations act through the characters $e^{i\langle q,\cdot\rangle}$, and $c_2$ enters only through the affine law~\eqref{eq:coadfinite}~\cite{Barnich:2014kra,Oblak:2017ptc}. The inducing data are representations of the preimage of $H_q$ in the Virasoro extension with central charge $c_1$. The next proposition shows that this extension does not affect the cyclic label.

\begin{proposition}\label{prop:central}
Every central extension of Lie groups $1\to\R\to E\to\R\times C_N\to1$ is abelian and split. Two splittings differ by a homomorphism $\R\times C_N\to\R$, which vanishes on $C_N$.
\end{proposition}

\begin{proof}
The preimage $E_0$ of $\R$ is a connected two dimensional Lie group whose Lie algebra has a one dimensional centre. The nonabelian two dimensional Lie algebra has trivial centre, so $E_0$ is abelian, and being a connected abelian extension of $\R$ by $\R$ it is $\R^2$. Let $\hat\gamma$ lift the generator of $C_N$. Then $\hat\gamma^N=c$ is central, and $\hat\gamma\cdot(-c/N)$ has order $N$. The map $\hat t\mapsto[\hat t,\hat\gamma]$ is a continuous homomorphism $E_0\to\R$ vanishing on the centre, because commutators are central, so it defines $\chi:\R\to\R$ with $N\chi=0$, since $\hat\gamma^N$ is central. Hence $\chi=0$, $E$ is abelian and $E\cong\R^2\times C_N$. A homomorphism from $C_N$ to $\R$ is trivial. This completes the proof.
\end{proof}

Thus the cyclic label is canonical, while the continuous label is defined up to a shift proportional to $c_1$. The Bott cocycle vanishes when one argument is a rigid rotation, so on the cyclic factor the splitting is even explicit. For Einstein gravity $c_1=0$, while theories with a gravitational Chern and Simons term, such as topologically massive gravity, have $c_1\neq0$. Proposition~\ref{prop:central} shows that the cyclic label and the spacing of the helicities are unaffected by $c_1$. Whether $c_1$ also shifts the helicities by a common constant is a separate question, which we leave open. It depends on the restriction of the Bott cocycle to the identity component of $\widetilde H_q$ and on the basis chosen for the Poincar\'e subalgebra: in the basis in which the Virasoro cocycle $\frac{c_1}{12}n(n^2-1)$ vanishes on $\sltwo$, the Poincar\'e subalgebra is not centrally extended, and the question is whether the splitting of Proposition~\ref{prop:central} can be chosen compatibly with it. For the massive particles of~\cite{Barnich:2014kra,Oblak:2017ptc} the central charge $c_1$ enters through the vacuum value of the rotation generator, and an analogous common shift on $\Om_{m,\Delta}$ is plausible, but we have not derived it. If such a shift is not a multiple of one half, the rotation through $2\pi$ acts by a phase, and the natural setting is the universal cover discussed below. The group $\R\times C_{2m}$ is abelian, so for $q$ invariant under $R_m$ in $\Om_{m,\Delta}$ the unitary irreducible representations of $H^{(2)}_q$ are the elements of its Pontryagin dual,
\begin{equation}\label{eq:chars}
\rho_{s,r}\big(\Phi_t\gamma^\ell\big)=e^{ist}\,e^{i\pi r\ell/m},\qquad s\in\R,\quad r\in\Z/2m\Z ,
\end{equation}
and on $G^{(N)}$ the cyclic label lies in $\Z/mN\Z$. The full little group is the semidirect product of $H^{(2)}_q$ with the supertranslations, on which the inducing representation is $(h,\alpha)\mapsto\rho_{s,r}(h)e^{i\langle q,\alpha\rangle}$. The label $s$ is conjugate to the flow of the hyperbolic superrotations preserving $q$. It plays the role of the spin of massive particles, but it is not quantized and has no Poincar\'e counterpart, because this flow never meets the Lorentz group. The label $r$ is the phase of a state under the finite rotational symmetry of the configuration.

On the universal cover the stabilizer is $\widetilde H_q\cong\R\times\Z$, with characters $e^{ist}e^{i\alpha j}$ on $\widetilde\Phi_t\widetilde R_m^{\,j}$ and $\alpha\in\R/2\pi\Z$. The rotation subgroup of the universal cover of $G_0$ is $\widetilde K\cong\R$, with $\widetilde k(\omega)$ translating by $2\omega$ and $\widetilde R_m=\widetilde k(\pi/m)$. The following proposition gives the corresponding weights.

\begin{proposition}\label{prop:anyon}
A locally square integrable function $f$ on $\widetilde K\cong\R$ with $f(\omega+\pi/m)=e^{-i\alpha}f(\omega)$ has Fourier support in $\frac{m\alpha}\pi+2m\Z$, and the corresponding helicities are $h=\frac{m\alpha}{2\pi}+mn$, $n\in\Z$.
\end{proposition}

\begin{proof}
The function $e^{i\alpha m\omega/\pi}f(\omega)$ has period $\pi/m$, so it equals $\sum_nc_ne^{-2imn\omega}$, and $f=\sum_nc_ne^{-i(m\alpha/\pi+2mn)\omega}$. Since $\widetilde k(\omega)$ rotates through $2\omega$, a weight $\sigma$ has helicity $\sigma/2$. This completes the proof.
\end{proof}

For $\alpha=\pi r/m$ this is the double cover ladder $h=r/2+mn$, and for $\alpha\in2\pi\Z/m$ the helicities are integers in one class modulo $m$. In general the fractional part of the helicity, and hence the statistics phase $e^{im\alpha}$, is common to the whole ladder, in agreement with the spin and statistics connection for anyons. The proposition extends to the whole global branch. The subgroup $\langle\widetilde R_m\rangle$ is cocompact in $\widetilde K$, and induction in stages gives
\[
\Ind_{\langle\widetilde R_m\rangle}^{\widetilde G_0}e^{i\alpha}\cong\bigoplus_{n\in\Z}\Ind_{\widetilde K}^{\widetilde G_0}\sigma_{m\alpha/\pi+2mn},
\]
where $\widetilde G_0$ is the universal cover of the Lorentz group and $\sigma_\sigma(\widetilde k(\omega))=e^{i\sigma\omega}$ for real $\sigma$. Each summand consists of the square integrable sections of a line bundle over $\Hb^2$ of real weight $\sigma$, the hyperbolic Landau problem with a magnetic charge that is no longer quantized. For $\sigma>1$ its discrete spectrum consists of the discrete series of $\widetilde G_0$ with lowest weights $\sigma-2j>1$, $j=0,1,\dots$, and similarly for $\sigma<-1$, so the analysis of section~\ref{sec:harmonic} carries over with real weights~\cite{Comtet:1986ki}. Under the smooth bundle assumptions of section~\ref{sec:topology}, the discrete phase $\alpha$ can be realized as flat holonomy when the character is trivial on the connected stabilizer. The restriction $\alpha=\pi r/m$ for the double cover is imposed by equivariance under $G^{(2)}$ rather than only under $\widetilde G$. This does not construct an invariant flat connection for a general nonzero continuous character $s$.

\subsection{Independence of the cyclic label}\label{sec:independence}

The characters~\eqref{eq:chars} can be chosen independently, but this does not yet show that $r$ is a new quantum number, since for other orbit types a similar phase might be fixed by the continuous data. Every orbit containing points invariant under $R_m$ gives an element $\gamma$ in their stabilizers. We compare how its action is determined in different orbit types. For a non exceptional constant $p_0$ the equation $2p_0u'-\kappa u'''=0$ has only constant periodic solutions, so the stabilizer in $G^{(2)}$ is the circle $\{k(\omega)\}$, with spin characters $\sigma_s(k(\omega))=e^{is\omega}$, $s\in\Z$, of spin $s/2$. The central result of this section is the following.

\begin{theorem}\label{thm:independence}
Let $m\ge2$, let $q$ be invariant under $R_m$, and let $\gamma\in H_q^{(2)}$ be the image of $\widetilde R_m$.
\begin{enumerate}[label=(\alph*),nosep]
\item If $q$ lies in $\Om_{n,\Delta}$ or in $T_{n,\pm}$ with $n\ge2$, let $\widetilde\delta$ be the unique element of $\widetilde H_q$ with $\widetilde\delta^{\,n}=T$, and $\gamma_n$ its image in $G^{(2)}$. Then $m\mid n$, $H^{(2)}_q\cong\R\times C_{2n}$ with $C_{2n}=\langle\gamma_n\rangle$, and $\gamma=\gamma_n^{n/m}$. Every character $\rho_{s,r_n}(\Phi^q_t\gamma_n^\ell)=e^{ist}e^{i\pi r_n\ell/n}$ satisfies $\rho_{s,r_n}(\gamma)=e^{i\pi r_n/m}$, independently of $s$.
\item If $q$ lies on the orbit of a non exceptional constant, then $H^{(2)}_q$ is a circle containing $\gamma$ and $\sigma_s(\gamma)=e^{i\pi s/m}$, so that $r\equiv s\pmod{2m}$ is fixed by the spin.
\item If $\gamma$ lies in the identity component of $H^{(2)}_q$, its action in any representation is determined by the restriction to the identity component.
\end{enumerate}
\end{theorem}

\begin{proof}
(a) By Theorem~\ref{thm:stabilizer} and Propositions~\ref{prop:hyperbolic} and~\ref{prop:parabolic}, $\widetilde H_q\cong Z(\widetilde M)\cong\R\times\Z$, with identity component $\R\times\{0\}$ and $T\mapsto(\lambda,n)$, where $\lambda=\pm1$ in the parabolic case. Since $\R\times\Z$ is torsion free, $n$th roots are unique, and $\widetilde\delta=(\lambda/n,1)$. Since $R_m\in H_q$, $\widetilde R_m=(\mu,j)$ with $\widetilde R_m^{\,m}=T$, so $m\mu=\lambda$ and $mj=n$. Hence $m\mid n$ and $\widetilde R_m=\widetilde\delta^{\,n/m}$. The homomorphism $(t,j)\mapsto(t-j\lambda/n,\ j\bmod 2n)$ maps $\R\times\Z$ onto $\R\times\Z_{2n}$ with kernel $\langle T^2\rangle$, so it identifies $H^{(2)}_q=\widetilde H_q/\langle T^2\rangle$ with $\R\times C_{2n}$ and maps $\widetilde\delta$ to $(0,1)$. Thus $\gamma_n$ generates the torsion factor, $\gamma=\gamma_n^{n/m}$, and $\rho_{s,r_n}(\gamma)=e^{i\pi r_n/m}$.
(b) Write $q=f\cdot p_0$. Then $f^{-1}R_mf\in K$ has the rotation number of $R_m$, so it equals $R_m$, and $\widetilde f\widetilde R_m\widetilde f^{-1}$ is the lift of $R_m$ with translation number $2\pi/m$, namely $\widetilde R_m$. Hence $H^{(2)}_q=\widetilde f\{k(\omega)\}\widetilde f^{-1}$ and $\gamma=\widetilde fk(\pi/m)\widetilde f^{-1}$, on which the transported $\sigma_s$ equals $e^{i\pi s/m}$.
(c) The identity component of a Lie group is generated by exponentials of its Lie algebra, and a continuous representation maps a product of exponentials to the product of the corresponding one parameter groups. This completes the proof.
\end{proof}

Finite cyclic global branches therefore exist on every orbit with points of finite rotational symmetry, including the massive particles of~\cite{Barnich:2014kra}, but there the phase of $\gamma$ is the spin phase. Only on the hyperbolic and parabolic orbits with winding is the cyclic factor a direct factor not contained in the identity component, so that its character is a new quantum number. For the exceptional constants $\gamma$ lies in the identity component of $\PSL^{(n)}$. It acts by one in the trivial representation, which for $n=1$ describes the vacuum, while the nontrivial unitary representations are infinite dimensional and contain several characters of $\langle\gamma\rangle$. Part (a) covers the parabolic orbits with winding as well. Moreover, the proof of Proposition~\ref{prop:family}(ii) shows that $\Lambda_m$ maps an orbit with lifted monodromy $\widetilde N$ into the orbit with lifted monodromy $\widetilde N^m$, and $(\exp(\pm Y)\zeta)^m=\exp(\pm mY)\zeta^{m}$ is conjugate to $\exp(\pm Y)\zeta^{m}$ by a diagonal matrix. The orbits $T_{1,\pm}$ are not empty. The element $\exp(\pm Y)\zeta$ is an increasing diffeomorphism of $\R$ without fixed points, so it is smoothly conjugate to the translation by $2\pi$: one prescribes an increasing $\theta$ on a fundamental interval with matching jets at its ends and extends it by $\theta(\varphi+2\pi)=\exp(\pm Y)\zeta\,\theta(\varphi)$. Lemma~\ref{lem:developing} then gives a periodic supermomentum with this lifted monodromy~\cite{Balog:1997zz}. Hence $\Lambda_m(T_{1,\pm})\subset T_{m,\pm}$ is a nonempty set of points invariant under $R_m$, and for them Theorem~\ref{thm:intersection} gives $G_0\cap H_q=\langle R_m\rangle$, since $R_m$ has order $m$ in the torsion subgroup $\Z_m$. With part (a) in place of Proposition~\ref{prop:covers}, the analysis of sections~\ref{sec:restriction} and~\ref{sec:twisted} therefore applies verbatim to these points, with the same finite shift equation.

The physical interpretation is the following. In a BMS$_3$ particle with constant supermomentum, a finite rotational symmetry of a dressed configuration is part of the rotational symmetry of the rest frame, and the phase under it is the spin phase. A configuration on $T_{m,\Delta}$ has no rest frame with continuous rotational symmetry. Its little group consists of hyperbolic superrotations and a finite rotation group, and the phase under the latter distinguishes quantum states built on the same classical configuration with the same $s$. Like a $\theta$ angle in gauge theory, it is invisible classically, a comparison made precise in section~\ref{sec:topology}.

%======================================================================
\section{Restriction to the global orbits}\label{sec:restriction}

We now relate the representations induced from $\Om_{m,\Delta}$ to the finite cyclic representations of $B(2,1)$. We set $c_1=0$, as in Einstein gravity. For $c_1\neq0$ the cyclic label is unchanged by Proposition~\ref{prop:central}, and a possible common shift of the helicities, left open in section~\ref{sec:chars}, would not affect the structure derived below.

\subsection{Induced representations and pullback}\label{sec:pullback}

We first recall the induced representations in a form that avoids measure theoretic issues and then restrict them to the Lorentz orbit through an invariant point. Fix $q\in\Om_{m,\Delta}$ invariant under $R_m$ and a character $\rho=\rho_{s,r}$. Following Wigner and Mackey~\cite{Wigner:1939cj,Mackey:1952ind} and~\cite{Barnich:2014kra,Oblak:2017ptc}, let
\begin{equation}\label{eq:smoothmodel}
\mathcal H^\infty_{s,r}=\Big\{\Psi\in C^\infty\big(G^{(2)}\big):\ \Psi(gh)=\rho(h)^{-1}\Psi(g),\ h\in H^{(2)}_q\Big\},
\end{equation}
where a function on the Fréchet Lie group $G^{(2)}$ is smooth if its composition with every smooth curve is smooth~\cite{Kriegl:1997}. We call~\eqref{eq:smoothmodel} the formal smooth induced model. It is not a Hilbert space, and no unitary structure on it is constructed in this paper. The operators~\eqref{eq:Uaction} below preserve it. Indeed, composition and inversion are smooth on $\Diff$ and its covers~\cite{Kriegl:1997}, so $g\mapsto g\cdot q$ is a smooth map into $C^\infty(S^1)$, and for each smooth supertranslation $\alpha$ the function $g\mapsto e^{i\langle g\cdot q,\alpha\rangle}$ is smooth, since the pairing is continuous and linear in its first argument. It is also invariant under right multiplication by $H^{(2)}_q$, so it preserves the equivariance condition, and left translations act smoothly. Thus~\eqref{eq:Uaction} defines an action of $G^{(2)}\ltimes\Vect$ on the formal smooth induced model, with smooth supertranslations. A superrotation $f$ and a supertranslation $\alpha$ act by
\begin{equation}\label{eq:Uaction}
\big(U(f,\alpha)\Psi\big)(g)=e^{i\langle g\cdot q,\alpha\rangle}\,\Psi\big(f^{-1}g\big).
\end{equation}
Heuristically, an element of the formal smooth induced model is a wavefunction on the orbit of Bondi mass aspects with values in the little group representation, supertranslations measure the supermomentum at each point, and superrotations move along the orbit. A unitary realization needs a quasi invariant measure on $\Om_{m,\Delta}$, whose construction is delicate~\cite{Oblak:2017ptc,Melas:2021hil}. No canonical bounded restriction map between the relevant $L^2$ spaces has been established, and we do not claim that the global branch occurs as a closed subrepresentation, or as a summand of positive measure in a direct integral, of a unitary BMS$_3$ representation. We therefore work with smooth sections, and we first specify the setting. The group $G^{(2)}$ is a Fréchet Lie group modelled on $\Vect$, and smoothness is understood in the convenient sense~\cite{Kriegl:1997}. In this section the supertranslations form the abelian Fréchet Lie group $\Vect\cong C^\infty(S^1)$, with its usual topology of uniform convergence of all derivatives, and the pairing $\langle p,\alpha\rangle=\int p\alpha\,d\varphi$ is a continuous bilinear form on $C^\infty(S^1)\times C^\infty(S^1)$. This smooth model must be distinguished from the Hilbert model of $B(2,1)$, in which the supertranslations form $L^2(S^1)$. For the unitary global branch the two are compatible, because $\beta=2\pi(q+\kappa/2)$ is smooth and therefore bounded, so $\alpha\mapsto\langle g_*\beta,\alpha\rangle_B$ extends continuously to $L^2(S^1)$, and the operators $V$ of Proposition~\ref{prop:dictionary} extend to the Hilbert model. No smoothness statement for the full orbit is transferred to the Hilbert model. The stabilizer $H^{(2)}_q\cong\R\times C_{2m}$ is a closed subgroup and a finite dimensional Lie group, by Theorem~\ref{thm:stabilizer} and the remark following it. The orbit $\Om_{m,\Delta}\cong G^{(2)}/H^{(2)}_q$ carries the quotient topology. Whether it is a smooth manifold for which $G^{(2)}\to\Om_{m,\Delta}$ is a locally trivial principal bundle is not proved here, and wherever this structure is used it is stated as an assumption. Precisely, the assumption is that $\Om_{m,\Delta}$ carries a structure of convenient manifold for which the projection is smooth and admits smooth local sections around every point, so that $G^{(2)}$ is locally diffeomorphic to the product of an open set of $\Om_{m,\Delta}$ with $H^{(2)}_q$. Let $\iota:\SL/C_{2m}\to\Om_{m,\Delta}$ be the orbit map of $q$, which is injective by Proposition~\ref{prop:covers}. If $\Om_{m,\Delta}$ has the assumed structure, $\iota$ is an injective immersion, because the infinitesimal action of $\sltwo$ at $q$ has kernel $\sltwo\cap\R u_q=0$: the stabilizer field $u_q$ has $2m\ge4$ zeros, while a nonzero element of $\sltwo$ has at most two. We do not need $\iota$ to be an embedding. The next theorem has two parts of different status. The statement about functions is unconditional. The identification of bundles holds as an $\SL$ equivariant bijection of homogeneous bundles over $\SL/C_{2m}$ in general, and as an isomorphism of smooth bundles whenever $\Om_{m,\Delta}$ has the assumed structure.

\begin{theorem}\label{thm:restriction}
The pullback $\iota^*\big(G^{(2)}\times_{H^{(2)}_q}\C_\rho\big)$ is isomorphic, equivariantly under $\SL$, to $\SL\times_{C_{2m}}\C_r$ with $\gamma\mapsto e^{i\pi r/m}$, independently of $s$. Restriction $\Psi\mapsto\Psi|_{\SL}$ maps $\mathcal H^\infty_{s,r}$ into
\begin{equation}\label{eq:Hglob}
\mathcal H^{\rm glob,\infty}_r=\Big\{\psi\in C^\infty\big(\SL\big):\ \psi(g\gamma)=e^{-i\pi r/m}\psi(g)\Big\}
\end{equation}
and intertwines the action of the subgroup $\SL\ltimes\Vect$ of Lorentz transformations and smooth supertranslations, a smooth version of $B(2,1)$, with $(U(f,\alpha)\psi)(g)=e^{i\langle g\cdot q,\alpha\rangle}\psi(f^{-1}g)$.
\end{theorem}

\begin{proof}
By Lemma~\ref{lem:pullback} the pullback is $\SL\times_{C_{2m}}\C_{\rho|_{C_{2m}}}$, and on $C_{2m}$ the flow parameter vanishes, so $\rho_{s,r}|_{C_{2m}}=\sigma_r$. For $\Psi\in\mathcal H^\infty_{s,r}$ and $g\in\SL$, $\Psi(g\gamma)=e^{-i\pi r/m}\Psi(g)$, and $\Psi|_{\SL}$ is smooth. For $f,g\in\SL$, $f^{-1}g\in\SL$, so~\eqref{eq:Uaction} on $\SL$ involves only values on $\SL$. This completes the proof.
\end{proof}

We make the restriction map precise. The space $\mathcal H^\infty_{s,r}$ carries the topology of $C^\infty(G^{(2)})$ in the convenient sense, and $\mathcal H^{\rm glob,\infty}_r$ the Fréchet topology of $C^\infty(\SL)$. Restriction is the pullback by the smooth inclusion $\SL\hookrightarrow G^{(2)}$, hence a bounded linear map, and it intertwines the action of $\SL\ltimes\Vect$ by Theorem~\ref{thm:restriction}. Smooth sections are needed because restriction to the three dimensional submanifold $\SL/C_{2m}$ is not defined on square integrable classes without further information on the measure. The next proposition shows that the restriction map is not trivial, under two explicit assumptions. The second one is a smooth regularity property, which is natural for manifolds modelled on nuclear Fréchet spaces~\cite{Kriegl:1997} but which we have not verified for $\Om_{m,\Delta}$. The pullback theorem itself does not depend on it.

\begin{proposition}\label{prop:nonzero}
Assume, as in section~\ref{sec:topology}, that $G^{(2)}\to\Om_{m,\Delta}$ is a locally trivial smooth principal bundle with structure group $H^{(2)}_q$, and that $\Om_{m,\Delta}$ admits smooth cutoffs: for every point $x_0$ and every open neighbourhood $U$ of $x_0$ there is a smooth function on $\Om_{m,\Delta}$ equal to one at $x_0$ whose support is closed and contained in $U$. Then for every $g_0\in\SL$ there is $\Psi\in\mathcal H^\infty_{s,r}$ with $\Psi(g_0)\neq0$. In particular the image of the restriction map is a nonzero subspace of $\mathcal H^{\rm glob,\infty}_r$, invariant under $\SL$, which contains for every point of $\SL$ a function not vanishing there.
\end{proposition}

\begin{proof}
Let $x_0=g_0\cdot q$, and choose a smooth local section $\sigma$ of the bundle over a neighbourhood $U$ of $x_0$. Replacing $\sigma$ by $\sigma h$ for a constant $h\in H^{(2)}_q$, we may assume $\sigma(x_0)=g_0$. By the second assumption there is a smooth $b$ on $\Om_{m,\Delta}$ with $b(x_0)=1$ and support closed and contained in $U$. Define $\Psi(\sigma(x)h)=\rho_{s,r}(h)^{-1}b(x)$ for $x\in U$ and $h\in H^{(2)}_q$, and $\Psi=0$ outside the preimage of $U$. In the trivialization $U\times H^{(2)}_q$ this is the product of $b$ with the smooth function $\rho_{s,r}^{-1}$, so $\Psi$ is smooth, and it vanishes near the boundary of the preimage of $U$. It satisfies $\Psi(gh)=\rho_{s,r}(h)^{-1}\Psi(g)$ and $\Psi(g_0)=b(x_0)=1$. The image is invariant under $\SL$ because restriction intertwines the left actions. This completes the proof.
\end{proof}

The map of Theorem~\ref{thm:restriction} is a restriction of formal smooth equivariant sections. It is not, by itself, a restriction operator on a unitary representation of the full infinite dimensional BMS$_3$ orbit. Proposition~\ref{prop:nonzero} is conditional on a smooth locally trivial quotient and the availability of smooth cutoffs, and these hypotheses have not been proved here. The conclusions of this paper do not depend on the density of the image. The finite shift equation, the helicity ladder and the twisted amplitudes are properties of $\mathcal H^{\rm glob}_r$ itself, and Theorem~\ref{thm:restriction} together with Proposition~\ref{prop:nonzero} shows that every restricted section obeys the finite shift equation and, under the hypotheses of Proposition~\ref{prop:nonzero}, that nonzero restricted sections exist. The helicity ladder and the twisted amplitudes are properties of the unitary global branch defined below. Density would be needed only for the converse statement that every vector of the global branch comes from a BMS$_3$ state.

We do not claim that this map is onto or has dense image. That would require extending smooth sections from the three dimensional orbit to the infinite dimensional one, for instance through a smooth slice built from the exponential map on a complement of $\sltwo\oplus\R u_q$ in $\Vect$, and the inverse function theorem is not available in general for Fréchet manifolds. Restrictions need not have compact support or finite norm, and Theorem~\ref{thm:restriction} does not guarantee either. We call $\mathcal H^{\rm glob}_r$, the Hilbert completion of the compactly supported elements of~\eqref{eq:Hglob} with respect to the norm $\|\psi\|^2=\int_{\SL/C_{2m}}|\psi|^2\,d\dot g=\frac1{2m}\int_{\SL}|\psi|^2\,dg$, the unitary global branch at $q$. Here $dg$ is the Haar measure of $\SL$ normalized in section~\ref{sec:explicit}, with $\mathrm{vol}(K)=2\pi$, and $d\dot g$ is the quotient measure on $\SL/C_{2m}$ obtained by giving each point of $C_{2m}$ mass $1/2m$, so that $\int_{\SL}f\,dg=2m\int_{\SL/C_{2m}}f\,d\dot g$ for every integrable $f$ invariant under right multiplication by $C_{2m}$. The integrand $|\psi|^2$ is such a function, because $|e^{i\pi r\ell/m}|=1$, so the first integral is defined and the two expressions agree. The same $dg$ is used in section~\ref{sec:twisted}, and with this convention the projections $P_r$ and the amplitudes $W_\ell$ refer to one and the same measure. Only the section $\psi$ itself, and not $|\psi|^2$, fails to descend to the quotient when $r\neq0$. Thus $\mathcal H^{\rm glob}_r$ is the space of square integrable equivariant functions on $\SL$, equivalently $L^2(\SL\times_{C_{2m}}\C_r)$, the square integrable sections of the homogeneous line bundle of Theorem~\ref{thm:restriction}. For $r\neq0$ it is not a space of functions on $\SL/C_{2m}$. It is a bona fide unitary representation of $\SL\ltimes\Vect$, and we study it independently of whether restrictions of elements of the formal smooth induced model are square integrable.

It is important to state what the global branch does and does not capture. Formally, if a quasi invariant measure on $\Om_{m,\Delta}$ exists, if the space of Lorentz orbits in $\Om_{m,\Delta}$ is a standard measurable space for the induced measure class, so that the measure disintegrates along the Lorentz orbits, and if the regularity hypotheses of Mackey's subgroup theorem hold, the restriction of the full representation to $\SL\ltimes\Vect$ is a direct integral over the Lorentz orbits $\SL\cdot q'\subset\Om_{m,\Delta}$ of the representations induced from $\SL\cap H^{(2)}_{q'}$~\cite{Mackey:1952ind}. By Theorem~\ref{thm:intersection}, $\SL\cap H^{(2)}_{q'}$ is finite cyclic, and it reduces to the centre $\{\pm I\}$ unless $q'$ is a Lorentz transform of a point invariant under some $R_{m'}$. For such $q'$ the summand is the subspace of $L^2(\SL)$ of the parity of $r$, which contains every cyclic sector of that parity. The exceptional points form a family of infinite codimension, but since no measure class on $\Om_{m,\Delta}$ has been specified we cannot say whether they form a null set. The global branch is the abstract induced representation associated with the Lorentz orbit of an exceptional point, and it must not be identified with a fibre of an unconstructed measure disintegration. No exceptional symmetric branch is identified here with a nonzero direct integral fibre. Its unitary representation is constructed independently. What is lost under restriction is therefore the dependence of a state on the infinitely many directions of $\Om_{m,\Delta}$ transverse to $\SL\cdot q$, which carries the remaining Fourier modes of the supermomentum and, through the generic fibres, the bulk of the Lorentz content. What survives is exactly the structure fixed by the little group: the label $r$, the finite shift equation, the helicity ladder and the twisted amplitudes of section~\ref{sec:twisted}. The label $r$ belongs to the inducing data, as the character of $H^{(2)}_q$, and the global branch is the place where it becomes visible to the Lorentz group. We have not shown that it is an observable superselection charge of quantum gravity. The finite functionals defined on the global branch, such as the twisted amplitudes, must not be identified with trace characters or physical observables of a full quantum BMS$_3$ theory.

We summarize the status of the results of this section. The following statements are unconditional: the restriction of smooth equivariant functions in Theorem~\ref{thm:restriction}, with its intertwining property, the identification of the restricted bundle as an $\SL$ equivariant bijection, the construction and unitarity of $\mathcal H^{\rm glob}_r$, and everything derived from $\mathcal H^{\rm glob}_r$ in section~\ref{sec:shift} and in sections~\ref{sec:poincare} and~\ref{sec:twisted}. The following statements depend on the assumption that $\Om_{m,\Delta}$ is a smooth quotient with smooth local sections: the identification of smooth bundles in Theorem~\ref{thm:restriction}, the immersion property of $\iota$, the computation of the fundamental group in Proposition~\ref{prop:pi1} and, together with the existence of smooth cutoffs, Proposition~\ref{prop:nonzero}. The existence of a quasi invariant measure, of a unitary representation on the full orbit and of its Mackey decomposition is not proved. No theorem of this paper uses it as a premise, and the remarks that refer to it are formulated conditionally.

The physical meaning of the restriction is simple. An element of the formal smooth induced model is a wavefunction on the space of gravitational configurations in $\Om_{m,\Delta}$. Its restriction records how the state depends on the configurations obtained from the symmetric configuration $q$ by Lorentz transformations, that is, by boosts and rotations of the asymptotic frame. Theorem~\ref{thm:restriction} shows that, along these directions, the only remnant of the full little group is the finite rotation group, and that the continuous label $s$ drops out completely. This is why the global branch is the natural place where the finite cyclic structure becomes visible to the Poincar\'e subgroup.

\subsection{Dictionary and finite shift equation}\label{sec:shift}

We now identify the global branch with a representation of $B(2,1)$ and solve its equivariance condition. For $g\in G_0$ in angular coordinates $\Sch(g)=\frac12(1-g'^2)$ (appendix~\ref{app:tech}), so~\eqref{eq:coadfinite} gives
\begin{equation}\label{eq:linearshift}
g\cdot p+\frac\kappa2=g_*\Big(p+\frac\kappa2\Big),\qquad g_*\varpi=\big(\varpi\circ g^{-1}\big)\big((g^{-1})'\big)^2 ,
\end{equation}
and the affine action of the Lorentz group becomes linear after the shift. In $B(2,1)$ supermomenta are quadratic densities paired with supertranslations by $\langle\beta,\alpha\rangle_B=\int\beta\alpha\,d\theta/2\pi$~\cite{Melas:2021hil,Melas:2026iwe}. The following proposition states the dictionary.

\begin{proposition}\label{prop:dictionary}
Identify $\theta$ with $\varphi$, set $\beta=2\pi(q+\kappa/2)$, and let $(V(f,\alpha)\psi)(g)=e^{i\langle g_*\beta,\alpha\rangle_B}\psi(f^{-1}g)$ be the $B(2,1)$ representation induced from $\beta$ and $\sigma_r$, on smooth supertranslations. Then $U(f,\alpha)=e^{-\frac{i\kappa}2\int\alpha\,d\varphi}V(f,\alpha)$ on $\mathcal H^{\rm glob,\infty}_r$, the little group of $\beta$ in $\SL$ is $C_{2m}$, and $U$ and $V$ have the same invariant subspaces.
\end{proposition}

\begin{proof}
By~\eqref{eq:linearshift}, $\langle g\cdot q,\alpha\rangle=\langle g_*\beta,\alpha\rangle_B-\frac\kappa2\int\alpha\,d\varphi$. The stabilizer of $\beta$ under the linear action equals that of $q$ under the affine one, intersected with $\SL$, which is $C_{2m}$. Scalar factors do not change invariant subspaces. This completes the proof.
\end{proof}

The realization of~\cite{Melas:2021hil,Melas:2026iwe} uses the right action $\upsilon\mapsto\upsilon g$ on the projective line, with $\upsilon=\cot(\theta/2)$, the Möbius map of $g^T$, so its dual action is the push forward by $g^{-T}=JgJ^{-1}$ with $J=k(\pi/2)$. The two realizations differ by this inner automorphism, a rotation of the circle through $\pi$ that fixes every $k(\omega)$ and hence $C_{2m}$ and $\sigma_r$. The Jacobian of $\upsilon\mapsto\upsilon g$ in the angular coordinate is the inverse conformal factor $\kappa_g^{-1}$ of~\cite{Melas:2021hil}, so its dual factor $\kappa_g^{-2}$ is the law of a quadratic density. The scalar factor in Proposition~\ref{prop:dictionary} is the contribution of the vacuum $-\kappa/2$, invisible to the Lorentz group.

Write $g=s(\xi)k(\omega)$ with $\xi\in\Hb^2$ and $s(\xi)$ positive definite, so that $dg=d\mu(\xi)\,d\omega$. Since $k(\omega)\gamma=k(\omega+\pi/m)$, the condition in~\eqref{eq:Hglob} is the finite shift equation
\begin{equation}\label{eq:fse}
\psi\Big(\xi,\omega+\frac\pi m\Big)=e^{-i\pi r/m}\,\psi(\xi,\omega).
\end{equation}
Its solutions are described by the following theorem.

\begin{theorem}\label{thm:floquet}
A measurable $\psi$ lies in $\mathcal H^{\rm glob}_r$ if and only if $\psi=\sum_n\psi_n(\xi)e^{-i(r+2mn)\omega}$ with $\sum_n\int|\psi_n|^2d\mu<\infty$, equivalently $\psi=e^{-ir\omega}F$ with $F$ of period $\pi/m$ in $\omega$. Right translations by $K$ commute with the left action and
\begin{equation}\label{eq:Kbranching}
\Ind_{C_{2m}}^{\SL}\sigma_r=\bigoplus_{n\in\Z}\Ind_K^{\SL}\sigma_{r+2mn},\qquad\sigma_{s'}(k(\omega))=e^{is'\omega},
\end{equation}
each summand occurring once.
\end{theorem}

\begin{proof}
For almost every $\xi$, $\omega\mapsto\psi(\xi,\omega)$ has a Fourier series $\sum_{s'}c_{s'}(\xi)e^{-is'\omega}$, and~\eqref{eq:fse} holds if and only if $c_{s'}=0$ for $s'\not\equiv r\pmod{2m}$. The norm on $\SL/C_{2m}$ is the integral of $|\psi|^2$ over $\Hb^2\times[0,\pi/m)$, which equals $\frac1{2m}\int_{\Hb^2}\int_0^{2\pi}|\psi|^2d\omega\,d\mu$, and Parseval gives the norm condition. Right translation by $k\in K$ commutes with $\gamma$ and preserves~\eqref{eq:fse}, and its eigenspaces are the spaces $\Ind_K^{\SL}\sigma_{s'}$ of functions with $\psi(gk(\omega))=e^{-is'\omega}\psi(g)$. This completes the proof.
\end{proof}

Equation~\eqref{eq:fse} is the finite shift equation derived intrinsically for $B(2,1)$ in~\cite{Melas:2026iwe} (appendix~\ref{app:B21}). Here it arises from the full nonconstant orbit, and $r$ is the monodromy of the bundle, not a Fourier coefficient of $q$. In contrast with a differential spin equation, it fixes the compact weight only modulo $2m$. When the projected momentum is nonzero and timelike, the right $K$ weights label the massive Poincar\'e helicities $r/2+mn$, by Theorem~\ref{thm:ladder}. Otherwise the same Fourier decomposition is purely representation theoretic. In either case the congruence states which weights are allowed. It does not imply that every allowed weight occurs with a nonzero coefficient in a given state.

%======================================================================
\section{Poincar\'e content and mass}\label{sec:poincare}

The restriction of a global branch to the Poincar\'e group gives the Poincar\'e content of the unitary global branch. We derive the helicity ladder, show that the Poincar\'e mass depends on the point of the orbit, and finally describe the branch under the Lorentz group alone.

\subsection{Helicity ladder}\label{sec:ladder}

In the centrally extended BMS$_3$ algebra the Poincar\'e generators close only after the vacuum value is subtracted from the zero mode, which on the global branch is the passage from $U$ to $V$ in Proposition~\ref{prop:dictionary}~\cite{Barnich:2014kra,Oblak:2017ptc}. We therefore restrict $V$ to the double cover $\mathcal P=\R^{2,1}\rtimes\SL$ of the Poincar\'e group, with $\R^{2,1}$ spanned by $f_0,f_1,f_2$. By Lemma~\ref{lem:projection} a translation $a$ acts by multiplication with $e^{ia^\mu(g\cdot P(q))_\mu}$. For $q$ invariant under $R_m$ with $m\ge2$ the function $q+\kappa/2$ has no modes $e^{\pm i\varphi}$, so $P(q)=(\mathsf M,0,0)$ with
\begin{equation}\label{eq:Mdef}
\mathsf M=P_0(q)=\int_0^{2\pi}\Big(q+\frac\kappa2\Big)d\varphi .
\end{equation}
For any point $q'$ the Lorentz invariant $C(q')=P_0(q')^2-P_1(q')^2-P_2(q')^2$ is constant on the Lorentz orbit of $q'$, by Lemma~\ref{lem:projection}. If $C>0$, the Lorentz orbit of $P(q')$ is a massive hyperboloid with mass $\sqrt C$ and a definite time orientation. If $C=0$, it is a null cone or the origin, and if $C<0$ it is spacelike and there is no rest energy. At an invariant point $P_1=P_2=0$, so $C(q)=\mathsf M^2\ge0$, and spacelike projections do not occur. For $\mathsf M\neq0$ the projection is timelike, $|\mathsf M|$ is the mass, and the sign of $\mathsf M$ selects the energy sheet. For $\mathsf M=0$ the projection vanishes and the massive analysis below does not apply. The component $P_0$ is frame dependent and equals $\mathsf M$ only in this rest frame. We call $\mathsf M$ the projected rest energy of $q$. For $\mu>0$, $\epsilon=\pm1$ and $h\in\frac12\Z$ let $U^{\rm Poin}_{\mu,\epsilon,h}$ be the irreducible massive representation of $\mathcal P$ with mass $\mu$, energy of sign $\epsilon$ and helicity $h$, induced from $\R^{2,1}\rtimes K$ and the character $e^{i\epsilon\mu a^0}e^{2ih\omega}$~\cite{Wigner:1939cj,Mackey:1952ind}. The ladder is described by the following theorem.

\begin{theorem}\label{thm:ladder}
Let $h_n=\frac r2+mn$ for $n\in\Z$.
\begin{enumerate}[label=(\roman*),nosep]
\item If $\mathsf M\neq0$, then $V|_{\mathcal P}\cong\bigoplus_{n\in\Z}U^{\rm Poin}_{|\mathsf M|,\epsilon,h_n}$ with $\epsilon$ the sign of $\mathsf M$. Every massive irreducible representation of mass $|\mathsf M|$, energy of sign $\epsilon$ and helicity $h_n$ occurs exactly once, and no other irreducible representation occurs.
\item If $\mathsf M=0$, the Poincar\'e translations act trivially, and $V|_{\mathcal P}$ is the representation $\mathcal H^{\rm glob}_r$ of $\SL$, extended trivially to $\R^{2,1}$. It is not a sum of massive representations, and its decomposition under $\SL$ is given by Proposition~\ref{prop:landau}.
\end{enumerate}
\end{theorem}

\begin{proof}
By Theorem~\ref{thm:floquet}, $\mathcal H^{\rm glob}_r=\bigoplus_n\Ind_K^{\SL}\sigma_{r+2mn}$. A translation $a\in\R^{2,1}$ acts by multiplication with $e^{ia^\mu(g\cdot P(q))_\mu}$, where $g\cdot P(q)$ is the Lorentz transform of $(\mathsf M,0,0)$. This function depends only on $gK$, because $K$ fixes $(\mathsf M,0,0)$, so the translations preserve each summand.
(i) Let $\mathsf M\neq0$. The vector $(\mathsf M,0,0)$ is timelike, its Lorentz orbit is the sheet of the hyperboloid of mass $|\mathsf M|$ with $\mathrm{sign}\,P_0=\epsilon$, and its stabilizer in $\SL$ is $K$. On the $n$th summand the representation of $\mathcal P$ is therefore induced from $\R^{2,1}\rtimes K$ and the character $e^{i\mathsf Ma^0}e^{i(r+2mn)\omega}$, which is $U^{\rm Poin}_{|\mathsf M|,\epsilon,h_n}$, since $k(\omega)$ rotates through $2\omega$ and the weight $r+2mn$ corresponds to the helicity $h_n$. By Wigner and Mackey~\cite{Wigner:1939cj,Mackey:1952ind} it is irreducible, and representations induced from the same orbit with different characters of $K$ are inequivalent. The summands have pairwise different weights $r+2mn$, which proves the multiplicity statement. For $\epsilon=-1$ the argument is the same, with the past sheet of the hyperboloid, on which $P_0\le\mathsf M<0$.
(ii) Let $\mathsf M=0$. Then $g\cdot P(q)=0$ for every $g\in\SL$, so every translation acts by the identity, and the representation of $\mathcal P$ is that of $\SL$ on $\mathcal H^{\rm glob}_r$. Its irreducible constituents under $\SL$ are discrete and principal series representations, each with infinite multiplicity, by Proposition~\ref{prop:landau}. As representations of $\mathcal P$ with trivial translations they are not massive. This completes the proof.
\end{proof}

This is the Poincar\'e decomposition of the independently constructed unitary global branch, not a decomposition of a full unitary BMS$_3$ representation on the entire infinite dimensional orbit. The representation of the global branch is defined even when no smooth restriction from a full orbit has been established.

The global branch at a point with timelike projection thus contains infinitely many massive Poincar\'e representations of the same mass, with helicities in an arithmetic progression. The step $m$ is fixed by the classical symmetry and the offset $r/2$ by the quantum monodromy. Even $r$ gives bosons and odd $r$ fermions, and on the universal cover the offset becomes $m\alpha/2\pi$. For massive BMS$_3$ particles the tower arises from the dressing of a rotation invariant configuration and no congruence beyond the spin constrains it~\cite{Barnich:2014kra}. Here the congruence class is the image of the monodromy. For $m=2$, for example, the sector $r=1$ contains only the helicities $\dots,-\frac72,-\frac32,\frac12,\frac52,\dots$, and the sector $r=2$ only $\dots,-3,-1,1,3,\dots$. The higher Fourier modes of the supermomentum, which are multiples of $m$ by the cyclic symmetry, are invisible to the Poincar\'e momentum, but they are responsible for the tower. Geometrically, the stabilizer of $(\mathsf M,0,0)$ in the Lorentz group is $K$, and the global orbit $\SL/C_{2m}$ fibres over the mass hyperboloid $\SL/K$ with fibre the circle $K/C_{2m}$. The finite shift equation is the condition on the dependence of a state along this fibre, and the helicity ladder is the Fourier decomposition along it. Through Proposition~\ref{prop:dictionary}, Theorem~\ref{thm:ladder} is the branching rule derived for the finite cyclic sectors of $B(2,1)$ in~\cite{Melas:2026iwe}, applied to the shifted supermomentum $\beta=2\pi(q+\kappa/2)$. What is new here is its origin. The ladder is not postulated by choosing a finite little group of the global group, but follows from the stabilizer of a gravitational configuration in the full superrotation group, and its offset is the monodromy of the corresponding bundle. The cases $\mathsf M\le0$ are equally relevant. At points with $\mathsf M<0$ the branch consists of massive representations of negative energy, and at points with $\mathsf M=0$ the translations act trivially and the branch is a representation of the Lorentz group alone, analysed in section~\ref{sec:harmonic}. The hypothesis $\mathsf M\neq0$ in Theorem~\ref{thm:ladder}(i) is therefore not automatic, and the value of $\mathsf M$ is the subject of the next subsection.

\subsection{Poincar\'e mass}\label{sec:mass}

For the representative the mass can be computed in closed form, as the following proposition shows.

\begin{proposition}\label{prop:massrep}
With $C=\cosh\eta$, $\frac1{2\pi}\int_0^{2\pi}p_{m,a}\,d\varphi=-\frac{\kappa m^2}2(4-2C-C^3)$ and
\begin{equation}\label{eq:massrep}
P_0(p_{m,a})=\pi\kappa\Big[1-m^2\big(4-2C-C^3\big)\Big].
\end{equation}
\end{proposition}

\begin{proof}
With $t=a\sin x$, $\frac{1+6t+4t^2}{(1+t)^2}=4-\frac2{1+t}-\frac1{(1+t)^2}$, and the averages of $(1+a\sin x)^{-1}$ and $(1+a\sin x)^{-2}$ are $C$ and $C^3$ (appendix~\ref{app:tech}). Averages are unchanged by $x=m\varphi$, and~\eqref{eq:Mdef} adds $\pi\kappa$. This completes the proof.
\end{proof}

As $\eta\to0$ the mass tends to $\pi\kappa(1-m^2)<0$, the value of the exceptional constant, so the projection is past directed for small $\eta$. It increases strictly with $\eta$ and vanishes at $\eta_*(m)$, where $m^2(4-2C-C^3)=1$. Numerically $\eta_*=0.5196,\,0.5606,\,0.5740,\,0.5801$ for $m=2,3,4,5$, with $\Delta_*=6.827,\,11.13,\,15.23,\,19.26$, and as $m\to\infty$ the threshold tends to the root $C=1.1795$ of $C^3+2C=4$. Without the central shift this would be the threshold for every $m$, which is incorrect.

These values refer to one point. By Proposition~\ref{prop:family} the invariant points are $\Lambda_m(w)$ with $w\in\Om_{1,\Delta/m}$, and
\begin{equation}\label{eq:P0lift}
P_0\big(\Lambda_m(w)\big)=m^2P_0(w)-\pi\kappa(m^2-1).
\end{equation}
The behaviour of $P_0$ on the orbit of $p_{1,a}$ is controlled by the full vector $P(p_{1,a})$, which we now compute.

\begin{proposition}\label{prop:sl2proj}
With $T=\tanh\eta$, $P(p_{1,a})=\big(\pi\kappa(C^3+2C-3),\,0,\,\kappa[\frac{2\pi}T(C-1)+\pi TC^3]\big)$. It is spacelike for $\eta<\eta_c$, lightlike at $\eta_c$ and future timelike for $\eta>\eta_c$, where $C_c=\cosh\eta_c=2.44240$ is the unique root larger than one of
\begin{equation}\label{eq:etac}
C^4-2C^2-6C-9=0 .
\end{equation}
Explicitly, $C_c=\frac12\big(s+\sqrt{2-2y_0+12/s}\big)$ with $s=\sqrt{2+2y_0}$, where
\[
y_0=-\frac13+\frac13\Big(\frac{26}{w}-w\Big),\qquad w=\Big(\frac{85+9\sqrt{3561}}4\Big)^{1/3},
\]
is the real root of $2y^3+2y^2+18y+9=0$.
\end{proposition}

\begin{proof}
$P_0$ is~\eqref{eq:massrep} with $m=1$, and $P_1=0$ because $\cos x\,(1+a\sin x)^{-k}$ is a derivative of a periodic function. $P_2$ follows from $\int\frac{\sin x\,dx}{1+a\sin x}=\frac{2\pi}a(1-C)$ and $\int\frac{\sin x\,dx}{(1+a\sin x)^2}=-2\pi aC^3$ (appendix~\ref{app:tech}). With $T^2=(C^2-1)/C^2$ one has $P_0=\pi\kappa(C-1)(C^2+C+3)$ and $P_2=\pi\kappa(C-1)(C^2+C+2)/T$, so that
\[
\frac{(P_2^2-P_0^2)(C^2-1)}{\pi^2\kappa^2(C-1)^2}=(C^2+C+2)^2C^2-(C^2+C+3)^2(C^2-1)=-C^4+2C^2+6C+9=:f(C),
\]
as one sees by writing $s=C^2+C$, for which the middle expression is $s^2+6s+9-(2s+5)C^2$. Now $f(1)=16$, $f''(C)=4-12C^2<0$ for $C\ge1$ and $f\to-\infty$, so $f$ has exactly one zero on $(1,\infty)$, where it changes sign from positive to negative. Since $P_0>0$ for $\eta>0$, the timelike vectors are future directed. For the explicit form we use Ferrari's method. Writing~\eqref{eq:etac} as $(C^2+y)^2=(2+2y)C^2+6C+9+y^2$, the right hand side is a perfect square exactly when $36=4(2+2y)(9+y^2)$, that is $2y^3+2y^2+18y+9=0$, whose only real root, the cubic being increasing, is $y_0\approx-0.51427$ by Cardano's formula. Then $2+2y_0>0$, and with $s=\sqrt{2+2y_0}$ the equation becomes $(C^2+y_0)^2=(sC+3/s)^2$. The factor $C^2-sC+y_0-3/s$ has the roots $\frac12\big(s\pm\sqrt{2-2y_0+12/s}\big)$, and the larger one is $C_c$, while the factor $C^2+sC+y_0+3/s$ has negative discriminant, since $s^2-4y_0-12/s<0$. This completes the proof.
\end{proof}

For the remaining case we need the following lemma, which gives a direct proof that the energy is unbounded below on the hyperbolic orbits with winding, in agreement with~\cite{Witten:1987ty,Balog:1997zz}.

\begin{lemma}\label{lem:unbounded}
Let $p$ be a supermomentum whose Hill equation $L_p\psi=0$ has a solution with two zeros at distance less than $2\pi$. Then $P_0$ takes every real value on $G\cdot p$. This applies to every point of $T_{n,\Delta}$ with $n\ge1$.
\end{lemma}

\begin{proof}
Let $g\in G$ have lift $f$ and $h=f^{-1}$. By~\eqref{eq:coadfinite}, $\int_0^{2\pi}g\cdot p\,d\varphi=\int_0^{2\pi}\big(p(h)h'^2-\kappa\Sch(h)\big)d\varphi$. Substituting $y=h(\varphi)$ and using $\Sch(h)=-(\Sch(f)\circ h)\,h'^2$, which follows from the cocycle identity applied to $f\circ h=\mathrm{id}$, gives $\int_0^{2\pi}(p+\kappa\Sch(f))f'^{-1}dy$. With $w=f'^{-1/2}>0$ one has $\Sch(f)=-2w''/w$, and an integration by parts gives
\[
\int_0^{2\pi}g\cdot p\,d\varphi=E[w]=\int_0^{2\pi}\big(2\kappa w'^2+pw^2\big)dy,\qquad\int_0^{2\pi}w^{-2}dy=2\pi .
\]
and $P_0(g\cdot p)=E[w]+\pi\kappa$. Conversely, every smooth periodic $w>0$ satisfying the constraint defines $f(y)=\int_0^yw^{-2}$, a lift of an element of $G$.

Let $\psi$ solve $L_p\psi=0$ with zeros $y_1<y_2$, $y_2-y_1<2\pi$. On an interval $I\supset[y_1,y_2]$ of length less than $2\pi$, the Dirichlet ground state energy of $-2\kappa\partial^2+p$ is negative. Indeed, on $[y_1,y_2]$ the restriction of $\psi$ is a Dirichlet eigenfunction of $-2\kappa\partial^2+p$ with eigenvalue zero, so the ground state energy there, being the smallest eigenvalue, is at most zero, also when $\psi$ has further zeros between $y_1$ and $y_2$. The ground state energy is the minimum of the quadratic form $Q[b]=\int(2\kappa b'^2+pb^2)$ over $b\in H^1_0$ with $\int b^2=1$. This is strict domain monotonicity, which we prove directly. Extending the ground state of $[y_1,y_2]$ by zero gives an admissible function on $I$, so the energy on $I$ is not larger. If it were equal, this extension would minimize the quadratic form on $I$ and hence be an eigenfunction on $I$ that vanishes on an open subinterval, which is impossible for a nonzero solution of a linear second order equation. Hence the ground state energy on $I$ is negative. The ground state on $I$ is positive inside $I$. We extend it by zero and mollify. Mollification preserves positivity, enlarges the support arbitrarily little and converges in $H^1$, and $Q$ is continuous on $H^1$. For a small mollification parameter we obtain a smooth $b\ge0$, supported in an interval of length less than $2\pi$, with $Q[b]<0$. Put $w_\lambda=c_\lambda(1+\lambda b)$, with $c_\lambda>0$ fixed by the constraint. As $\lambda\to\infty$, $c_\lambda^2=\frac1{2\pi}\int(1+\lambda b)^{-2}$ tends to $\frac1{2\pi}$ times the measure of $\{b=0\}$, which is positive, and $E[w_\lambda]=c_\lambda^2\big(\lambda^2Q[b]+2\lambda\int pb+\int p\big)\to-\infty$. Choosing instead $b$ with $Q[b]>0$, for instance a bump function with large derivative, gives $E[w_\lambda]\to+\infty$. Since $G$ is connected and $g\mapsto P_0(g\cdot p)$ is continuous, $P_0$ takes every real value on $G\cdot p$.

For $p\in T_{n,\Delta}$ we may take $\widetilde M=\exp(\lambda X)\zeta^n$ with $n\ge1$ by Proposition~\ref{prop:hyperbolic}. The fixed points of the hyperbolic flow $\exp(tX)$ on the line of angles divide it into intervals on which the flow moves points alternately up and down, so there is $y$ with $\exp(\lambda X)y>y$. Put $x=y-n\pi$. Then $\widetilde Mx=\exp(\lambda X)y>y\ge x+\pi$. The solution $\cos(\theta-x+\frac\pi2)/\sqrt{\theta'}$ vanishes at $\varphi_1=\theta^{-1}(x)$ and $\varphi_2=\theta^{-1}(x+\pi)$, and $\theta(\varphi_1+2\pi)=\widetilde Mx>\theta(\varphi_2)$ gives $\varphi_2-\varphi_1<2\pi$. This completes the proof.
\end{proof}

The projected rest energy is therefore not an invariant of the orbit, as the following theorem shows.

\begin{theorem}\label{thm:massrange}
The projected rest energy $\mathsf M$ at the cyclically invariant points $\Lambda_m(\Om_{1,\Delta/m})$ ranges over $\R$. At the points with $\mathsf M\neq0$ the projected momentum is timelike, with mass $|\mathsf M|$ and time orientation given by the sign of $\mathsf M$. For $\eta<\eta_c$ the statement follows from the Lorentz group alone.
\end{theorem}

\begin{proof}
For $g\in G_0$, $P(g\cdot p_{1,a})$ is the Lorentz transform of $P(p_{1,a})$, and $\Lambda_m(g\cdot p_{1,a})=\hat g\cdot p_{m,a}$ lies in the family. If $P(p_{1,a})$ is spacelike, boosts give every value of $P_0$, and by~\eqref{eq:P0lift} so on the family. In general, Lemma~\ref{lem:unbounded} applied to $p_{1,a}\in\Om_{1,\Delta/m}=T_{1,\Delta/m}$ shows that $P_0$ takes every real value on $\Om_{1,\Delta/m}$, hence by~\eqref{eq:P0lift} on the family, where $\mathsf M=P_0$. This completes the proof.
\end{proof}

Theorem~\ref{thm:massrange} is a statement about many different Lorentz orbits inside $\Om_{m,\Delta}$. On each of them the invariant $C=\mathsf M^2$ is constant, and no Casimir varies within a single global branch. This does not contradict the fact that the mass is a Casimir invariant of the Poincar\'e group. By Lemma~\ref{lem:projection}, $P_0^2-P_1^2-P_2^2$ is invariant under $G_0$, so it is constant on each Lorentz orbit inside $\Om_{m,\Delta}$. For a fixed Poincar\'e subgroup, two points of the family with different masses lie on different Lorentz orbits, and the superrotation relating them lies outside $G_0$ and does not normalize the Poincar\'e subgroup. Equivalently, transporting one point to the other by this superrotation conjugates the embedding of the Poincar\'e subgroup in BMS$_3$. The square of the mass is a Casimir of the Poincar\'e algebra, not of the BMS$_3$ algebra.

The invariant content of Theorem~\ref{thm:ladder} is thus the helicity ladder, not the mass. Unlike the massive BMS$_3$ particles, whose orbits have a minimum of the energy, the representations induced from $\Om_{m,\Delta}$ are not of positive energy and should not be read as stable one particle states. This does not affect unitarity. Conditionally, if a quasi invariant measure $\mu$ on the full orbit and a corresponding unitary induced representation with the required measurable structure can be constructed, the supertranslations would act on the associated $L^2$ space by multiplication with phases. The energy is the self adjoint generator of time translations, acting by multiplication with $P_0(q')$, and its spectrum is the essential range of $P_0$ with respect to $\mu$. By Lemma~\ref{lem:unbounded}, $P_0$ is continuous and takes every real value on the orbit. Under the additional assumption that $\mu$ has full support in the orbit topology and the energy generator is realized as this multiplication operator, its spectrum would be the whole real line. This spectral conclusion is conditional and is distinct from the unconditional classical range statement of Theorem~\ref{thm:massrange}. If such a unitary realization exists, it would resemble the tachyonic representations of the Poincar\'e group in allowing unbounded energy without contradicting unitarity.

This statement concerns the full orbit and must not be transferred to a single global branch. On the branch at an invariant point $q$, the Hilbert space structure is explicit: $\mathcal H^{\rm glob}_r=L^2(\SL\times_{C_{2m}}\C_r)$, realized as square integrable equivariant functions on $\SL$ with the norm of section~\ref{sec:pullback} and the Haar measure normalized in section~\ref{sec:explicit}, and $V$ is manifestly unitary, the supertranslations acting by the phases $e^{i\langle g_*\beta,\alpha\rangle_B}$ of modulus one. The energy of $V$ acts by multiplication with $P_0(g\cdot q)$, where $P(g\cdot q)$ is the Lorentz transform of $(\mathsf M,0,0)$. Its spectrum is therefore $[\mathsf M,\infty)$ for $\mathsf M>0$, $(-\infty,\mathsf M]$ for $\mathsf M<0$, and $\{0\}$ for $\mathsf M=0$. Each massive branch has energy of a definite sign, bounded away from zero, and only the family of all invariant points, or the full orbit, carries energies of both signs. The original operators $U$ differ from $V$ by the factor $e^{-\frac{i\kappa}2\int\alpha\,d\varphi}$ of Proposition~\ref{prop:dictionary}. For a time translation $\alpha=a^0$ this factor is $e^{-i\pi\kappa a^0}$, so the energy of $U$ is that of $V$ shifted by the constant $-\pi\kappa$, the energy measured from $p=0$ rather than from the vacuum. It remains an open question whether these inducing characters lead to distinct sectors in a full quantum theory or enter a gravitational path integral. No state space or physical observable implementing such a superselection interpretation has been constructed here. Maxfield's exact path integrals over $T_{n,\Delta}$~\cite{Maxfield:2026men} show that these orbits are legitimate integration domains, and the winding cosmologies are classical solutions with these orbits as phase spaces, so they are natural candidates for saddle contributions in a sum over orbits. We do not attempt to decide here whether such contributions are present in flat space quantum gravity.

\subsection{Harmonic analysis on the global orbit}\label{sec:harmonic}

The decomposition~\eqref{eq:Kbranching} reduces the analysis of a global branch under the Lorentz group alone to homogeneous line bundles over the hyperbolic plane. For $s'\in\Z$ let $\mathcal L_{s'}=\Ind_K^{\SL}\sigma_{s'}$ be the space of square integrable sections of the line bundle of weight $s'$ over $\Hb^2=\SL/K$. Its spectral decomposition is the hyperbolic Landau problem~\cite{Comtet:1986ki}. We use the convention $\psi(gk)=\sigma_{s'}(k)^{-1}\psi(g)$ of~\eqref{eq:smoothmodel}, for which Frobenius reciprocity states that an irreducible representation occurs in $\mathcal L_{s'}$ if and only if it contains the $K$ type $s'$. Recall that the discrete series $D^+_k$ with $k\ge2$ has $K$ types $k,k+2,\dots$ and $D^-_k$ has $K$ types $-k,-k-2,\dots$~\cite{Knapp:1986}. The structure of the branch is described by the following proposition.

\begin{proposition}\label{prop:landau}
The representation $\mathcal L_{s'}$ decomposes as
\begin{equation}\label{eq:landau}
\mathcal L_{s'}\cong\bigoplus_{\substack{2\le k\le|s'|\\ k\equiv s'\,(2)}}D^{\mathrm{sgn}\,s'}_k\ \oplus\ \int^\oplus_{\nu>0}\pi_{\nu,\epsilon}\,d\mu_\epsilon(\nu),\qquad\epsilon\equiv s'\pmod2 ,
\end{equation}
where $\pi_{\nu,\epsilon}$ are the unitary principal series of parity $\epsilon$ and $\mu_\epsilon$ is the Plancherel measure, equivalent to Lebesgue measure on $(0,\infty)$, each representation occurring once. Consequently, in the global branch $\Ind_{C_{2m}}^{\SL}\sigma_r$ each discrete series $D^\pm_k$ with $k\equiv r\pmod2$ occurs with multiplicity equal to the number of $s'\equiv r\pmod{2m}$ with $\pm s'\ge k$, which is infinite, and each principal series of parity $r\bmod2$ occurs with infinite multiplicity.
\end{proposition}

\begin{proof}
By Frobenius reciprocity for square integrable representations, the multiplicity of a discrete series representation $D$ in $\mathcal L_{s'}$ equals the multiplicity of the $K$ type $s'$ in $D$, which is one if $s'$ is a $K$ type of $D$ and zero otherwise~\cite{Knapp:1986}. This gives the discrete part. By the Plancherel theorem for line bundles over $\Hb^2$, every principal series representation of the parity of $s'$ contains the $K$ type $s'$ exactly once, which gives the continuous part~\cite{Knapp:1986,Comtet:1986ki}. For $\epsilon=1$ the representation at $\nu=0$ is reducible, which is irrelevant because it carries measure zero. Summing over $s'\equiv r\pmod{2m}$ with~\eqref{eq:Kbranching}, the set of $s'$ with $s'\equiv r\pmod{2m}$ and $\pm s'\ge k$ is infinite for every $k$, and every principal series of parity $r$ occurs in each summand. This completes the proof.
\end{proof}

The global branch is therefore not an admissible representation of $\SL$, since every $K$ type of parity $r$ occurs with infinite multiplicity. If the operators $\int\phi(g)V(g)\,dg$ were of trace class for all test functions $\phi$, every discrete series would occur with finite multiplicity, because its character is a nonzero distribution. Hence the branch has no distribution character, and its spectral information must be extracted from finite diagonal amplitudes, which is done in section~\ref{sec:twisted}. The label $r$ enters the discrete spectrum through the parity condition $k\equiv r\pmod2$ and through the multiplicity spaces, which are spanned by the weights in the class of $r$ modulo $2m$.

The physical interpretation is that of a charged particle on the mass hyperboloid. The weight $s'$ acts as a magnetic charge on $\Hb^2$, the discrete series are the Landau levels, whose number $\lfloor|s'|/2\rfloor$ grows with the charge, and the principal series are the scattering states. A global branch contains all charges in the class of $r$ modulo $2m$. In the language of Theorem~\ref{thm:ladder} these charges are twice the helicities, so the Landau levels are the bound states of the boost dynamics of a particle of helicity $h_n$ on its mass shell.

%======================================================================
\section{Bulk geometry and soft hair}\label{sec:bulk}

Supermomentum orbits label families of locally flat Bondi solutions with prescribed asymptotic data. We exhibit solutions associated with $\Om_{m,\Delta}$, establish the existence of two commuting Killing fields and compute their Lorentz holonomy. We then prove an obstruction in the smooth real periodic near horizon current model; neither a classification of global completions nor a no go theorem for all possible near horizon descriptions is claimed.

\subsection{Winding cosmologies}\label{sec:bondi}

In Bondi gauge every solution with the boundary conditions of~\cite{Barnich:2006av} reads~\cite{Barnich:2006av,Barnich:2012aw}
\begin{equation}\label{eq:bondi}
ds^2=\Theta\,du^2-2\,du\,dr+2\Big(\Xi+\frac u2\Theta'\Big)du\,d\varphi+r^2d\varphi^2 ,
\end{equation}
with periodic $\Theta(\varphi)$ and $\Xi(\varphi)$. With $f=T+uY'$ and $\mathcal N=\Xi+\frac u2\Theta'$ the asymptotic symmetries are the vector fields
\begin{equation}\label{eq:xi}
\xi_{(Y,T)}=f\,\partial_u+\Big(-rY'+f''-\frac1r\mathcal Nf'\Big)\partial_r+\Big(Y-\frac1rf'\Big)\partial_\varphi
\end{equation}
of~\cite{Barnich:2012aw}, whose radial component depends on the solution, and they act by
\begin{equation}\label{eq:thetaxi}
\delta\Theta=Y\Theta'+2Y'\Theta-2Y''',\qquad\delta\Xi=Y\Xi'+2Y'\Xi+\tfrac12\big(T\Theta'+2T'\Theta-2T'''\big),
\end{equation}
up to an overall sign convention~\cite{Barnich:2006av,Barnich:2012aw,Oblak:2017ptc}. The factor $\frac12$ can be checked directly: $(Y,T)=(0,1)$ gives $\partial_u$, which shifts $\Xi$ by $\frac12\Theta'$ because the metric depends on $u$ only through $\mathcal N$. Comparison with~\eqref{eq:coadfinite} gives $\Theta=2p/\kappa$, so Minkowski space is $\Theta=-1$, conical defects are constants $-1<\Theta<0$, and flat space cosmologies, the flat limits of BTZ black holes~\cite{Barnich:2012aw}, are constants $\Theta>0$. The energy measured from the vacuum is $\frac1{16\pi G}\int(\Theta+1)\,d\varphi$~\cite{Barnich:2012aw}, in agreement with Lemma~\ref{lem:projection}. For our orbit we take $\Theta_{m,a}=2p_{m,a}/\kappa$ and smooth periodic $\Xi$. The isometries are given by the next proposition.

\begin{proposition}\label{prop:killing}
Let $u=u_{m,a}$ and $\mathcal L_\Theta T=T\Theta'+2T'\Theta-2T'''$. There is a smooth periodic $T_\Xi$ with $\frac12\mathcal L_{\Theta_{m,a}}T_\Xi=-(u\Xi'+2u'\Xi)$, and $\xi_{(0,u)}$ and $\xi_{(u,T_\Xi)}$ are commuting Killing vectors of~\eqref{eq:bondi} with $\Theta=\Theta_{m,a}$. For $\Xi=0$ one may take $T_\Xi=0$, and if $\Xi$ is invariant under $R_m$, then $R_m$ is an isometry.
\end{proposition}

\begin{proof}
The vector fields~\eqref{eq:xi} preserve the gauge, and $\mathcal L_\xi g$ equals the variation of~\eqref{eq:bondi}, which is affine in $(\Theta,\Xi)$, under~\eqref{eq:thetaxi}, exactly and not only to leading order in $1/r$~\cite{Barnich:2006av,Barnich:2012aw}. Hence $(Y,T)$ with $\delta\Theta=\delta\Xi=0$ gives a Killing vector. For $(0,u)$, $\delta\Theta=0$ and $\delta\Xi=\frac1\kappa(up'+2u'p-\kappa u''')=0$. For $(u,T)$, $\delta\Theta=0$, and $\delta\Xi=0$ is the stated equation. The operator $\mathcal L_\Theta$ is skew adjoint for the $L^2$ pairing, since $X\mathcal L_\Theta T+T\mathcal L_\Theta X$ is a total derivative, and as an elliptic operator of order three on the circle it is Fredholm with range orthogonal to its kernel. The kernel consists of the stabilizer fields of $\Theta$, the line spanned by $u$. The right hand side is orthogonal to $u$, because $u(u\Xi'+2u'\Xi)=(u^2\Xi)'$, so a smooth solution exists. The fields~\eqref{eq:xi} depend on the solution, so their Lie bracket is the field of the BMS$_3$ bracket up to terms coming from the variation of $\mathcal N$~\cite{Barnich:2012aw}, which vanish because both fields leave $(\Theta,\Xi)$ invariant. The BMS$_3$ bracket $[(0,u),(u,T_\Xi)]$ vanishes, its supertranslation part being $u'u-uu'$. The last statement is clear. This completes the proof.
\end{proof}

The two displayed Killing fields span a two dimensional abelian Lie algebra wherever they are linearly independent, as happens for the standard flat space cosmologies, whose isometries $\partial_u$ and $\partial_\varphi$ correspond to $(0,1)$ and $(1,0)$. Here, however, the generating function has $2m$ zeros. At null infinity the Killing field $u_{m,a}\partial_u$ vanishes on $2m$ null generators, which divide null infinity into wedges in which the Killing time runs alternately forward and backward, and $R_m$ shifts the wedges by two. By Theorem~\ref{thm:massrange} the energy takes arbitrarily negative values on the orbit, while for flat space cosmologies the constant representative minimizes it.

Three dimensional gravity is a Chern and Simons theory for the Poincar\'e group~\cite{Achucarro:1986uwr,Witten:1988hc}, and locally flat solutions are characterized by their holonomies. The angular component of the Lorentz part of the gauge field is gauge equivalent to
\begin{equation}\label{eq:aphi}
a_\varphi=\begin{pmatrix}0&1\\ \Theta/4&0\end{pmatrix}
\end{equation}
in the fundamental representation~\cite{Barnich:2012aw,Oblak:2017ptc}. In a basis with $[L_a,L_b]=(a-b)L_{a+b}$, for instance $L_1=-E_{21}$, $L_{-1}=E_{12}$, $L_0=\frac12\mathrm{diag}(1,-1)$, this is $L_{-1}-\frac\Theta4L_1$. The relative sign matters, since the opposite sign would make the holonomy of flat space cosmologies elliptic. The following lemma connects the holonomy with section~\ref{sec:hill}.

\begin{lemma}\label{lem:holonomy}
The holonomy of~\eqref{eq:aphi} is conjugate to the monodromy of $\psi''=\frac\Theta4\psi$, which for $\Theta=2p/\kappa$ is the monodromy of $L_p$.
\end{lemma}

\begin{proof}
A parallel section satisfies $y_1'=y_2$ and $y_2'=\frac\Theta4y_1$, so its first component solves the Hill equation, and the transport matrix is a Wronskian matrix of solutions, defined up to conjugation. This completes the proof.
\end{proof}

By Proposition~\ref{prop:floquet} the Lorentz holonomy has trace $(-1)^m2\cosh(\Delta/2)$, and under $\SL\to SO^+(2,1)$ it is a boost of rapidity $\Delta$, whose lift to $\tSL$ has translation number $m\pi$. A flat space cosmology with constant $\Theta_0$ has Floquet multipliers $e^{\pm\pi\sqrt{\Theta_0}}$, a boost of rapidity $2\pi\sqrt{\Theta_0}$ and translation number zero. The two holonomies are conjugate in $SO^+(2,1)$ if and only if $\Theta_0=m^2\sinh^2\eta$, but by Corollary~\ref{cor:classification} the solutions are not related by any asymptotic symmetry. We call the corresponding Bondi solutions winding cosmology-type BMS geometries. This terminology refers to local flatness, hyperbolic Lorentz holonomy and the boundary winding class, and does not assert geodesic completeness, a regular event horizon or global equivalence to an ordinary flat space cosmology. They are the flat counterparts of the nonconstant geometries in anti de Sitter space classified in~\cite{Garbarz:2014kaa}. Geometrically, the developing map modulo $\pi$ maps the celestial circle to the circle of null directions. For the vacuum $\theta=\varphi/2$ wraps it once, in agreement with the holonomy $-I$, for a flat space cosmology it does not wrap it, and for a winding cosmology it wraps it $m$ times while the holonomy is a boost. At an $R_m$ invariant point, the rotation is a discrete symmetry of the boundary data and $r$ is the associated inducing character; interpreting it as a phase of a physical gravitational quantum state awaits construction of the full state space. The continuous character $s$ labels the corresponding stabilizer flow.

\subsection{Near horizon soft hair}\label{sec:softhair}

The winding geometries share their hyperbolic Lorentz holonomy with ordinary flat space cosmologies, and it is natural to examine the particular smooth real near horizon current realization of~\cite{Afshar:2016kjj}. Afshar, Grumiller, Merbis, Perez, Tempo and Troncoso showed that flat space cosmologies with soft hair are described near the horizon by abelian current algebras. In~\cite{Afshar:2016kjj} the asymptotic charges are composites of near horizon currents $(\omega,\gamma)$,
\begin{equation}\label{eq:miura}
\mathcal M=\tfrac12\omega^2+\omega',\qquad\mathcal L=\omega\gamma+\gamma'.
\end{equation}
If $\delta\omega=(Y\omega)'-Y''$, then $\delta\mathcal M=Y\mathcal M'+2Y'\mathcal M-Y'''$, so the Hill operator of $\mathcal M$ is $\psi''=\frac{\mathcal M}2\psi$. This does not depend on normalization: if $\Theta=\lambda\mathcal M$, then $\Theta$ has anomaly coefficient $\lambda$ and its Hill operator $\psi''=\frac\Theta{2\lambda}\psi$ is the same. Only the coadjoint transformation law of $\mathcal M$ matters, and it fixes the coefficient of $Y''$ in $\delta\omega$. The following theorem shows that the image of the Miura map misses all orbits with winding.

\begin{theorem}\label{thm:nogo}
If $\omega$ is real, smooth and periodic, then every solution of $\psi''=\frac{\mathcal M}2\psi$ has at most one zero on $\R$, and the lifted monodromy has translation number zero. In particular the supermomentum $\kappa\mathcal M$ does not belong to $\Om_{n,\Delta}$ for any $n\ge1$.
\end{theorem}

\begin{proof}
The function $\psi_0=\exp(\frac12\int_0^\varphi\omega)>0$ satisfies $\psi_0''=(\frac14\omega^2+\frac12\omega')\psi_0=\frac{\mathcal M}2\psi_0$. By the Sturm separation theorem (appendix~\ref{app:tech}), any solution independent of $\psi_0$ has a zero of $\psi_0$ between two of its zeros, hence at most one zero. By the zero count of section~\ref{sec:coadjoint}, $\tau(\widetilde M)=0$, while on $\Om_{n,\Delta}$ solutions have $n$ zeros per period. This completes the proof.
\end{proof}

The image of the current map thus lies in the topologically trivial sector, which contains the flat space cosmologies. Minkowski space and the conical defects are not in the image either, since their Hill equations have oscillating solutions. This is consistent with the original application to flat space cosmologies. The obstruction concerns smooth real periodic currents in the Miura model~\eqref{eq:miura}; it does not exclude complex, distributional or singular current descriptions, nor all conceivable near horizon boundary conditions.

A weaker statement survives. If $\omega$ and $\gamma$ have Fourier support in $m\Z$, so do $\mathcal M$ and $\mathcal L$, because differentiation preserves Fourier support and the support of a product lies in the sum of the supports. This $m$ graded sector of the soft hair populates the invariant points $\Lambda_m(w)$ of trivial orbits, where by Theorem~\ref{thm:independence}(b) the cyclic label is fixed by the spin, so the grading carries no information beyond the classical periodicity. The converse fails, since $\omega\circ R_m$ is again a periodic solution of the Riccati equation $\frac12\omega^2+\omega'=\mathcal M$, which in general differs from $\omega$, so an $m$ periodic composite may come from a current that is not $m$ periodic.

Physically, the near horizon description assumes a horizon generated by a Killing field without zeros, whereas the winding cosmologies have a Killing field with $2m$ zeros at null infinity. A current description of the winding cosmologies is nevertheless possible if singular currents are admitted. Let $\mathcal M=p_{m,a}/\kappa$, so that $\psi''=\frac{\mathcal M}2\psi$ is the Hill equation of $p_{m,a}$, and let $\psi_1$ be the Floquet solution of Proposition~\ref{prop:floquet}. The current $\omega=2\psi_1'/\psi_1$ is periodic, because $\psi_1$ is multiplied by a constant over one period, it is smooth away from the $m$ zeros of $\psi_1$ per period, where it has simple poles with residue $2$, and it satisfies $\frac12\omega^2+\omega'=2\psi_1''/\psi_1=\mathcal M$. For this explicit Floquet solution the number of poles per period equals the winding label. This realizes the winding information in a singular Riccati current, but does not by itself define an admissible smooth near horizon phase space.

Singular currents raise the question of whether the transformation law and the charges remain meaningful. The next proposition shows that the poles are carried along consistently and that the zero mode keeps its smooth meaning, provided it is defined by a principal value.

\begin{proposition}\label{prop:singular}
Let $\omega$ be periodic and smooth except for simple poles of residue $2$.
\begin{enumerate}[label=(\roman*),nosep]
\item Under $\delta\omega=(Y\omega)'-Y''$ the singular part at a pole $\varphi_0$ changes only by $-2Y(\varphi_0)(\varphi-\varphi_0)^{-2}$, so the pole moves with $\delta\varphi_0=-Y(\varphi_0)$ and keeps its residue.
\item If $\psi=\exp(\frac12\int\omega)$, defined up to sign across the poles, solves $\psi''=\frac{\mathcal M}2\psi$ with multiplier $\mu$ over one period, then the principal value of $\oint\omega$ equals $2\log|\mu|$. For the current $\omega=2\psi_1'/\psi_1$ of $p_{m,a}$ it equals $\Delta$, and for every smooth current it equals the signed logarithmic Floquet exponent $2\log\mu$; when the Lorentz monodromy is hyperbolic, its nonnegative boost rapidity is $2|\log\mu|$.
\end{enumerate}
\end{proposition}

\begin{proof}
(i) Write $\omega=2t^{-1}+c_0+c_1t+O(t^2)$ and $Y=\sum_jy_jt^j$ with $t=\varphi-\varphi_0$. Then $(Y\omega)'-Y''=-2y_0t^{-2}+O(1)$, because the terms $2y_1t^{-1}$ from $Y'\omega$ and $-2y_1t^{-1}$ from $Y\omega'$ cancel. The variation of $2(\varphi-\varphi_0)^{-1}$ under $\varphi_0\mapsto\varphi_0+\delta\varphi_0$ is $2\delta\varphi_0\,t^{-2}$, which gives $\delta\varphi_0=-y_0$.
(ii) Away from the poles $\omega=2(\log|\psi|)'$, and $\psi$ has simple zeros at the poles. Near such a zero $\log|\psi|=\log|t|+O(1)$, so excising symmetric intervals around the zeros and letting their length tend to zero gives $\mathrm{PV}\oint\omega=2\log|\psi(\varphi+2\pi)/\psi(\varphi)|=2\log|\mu|$. For $\psi_1$, Proposition~\ref{prop:floquet} gives $|\mu|=e^{\Delta/2}$. A smooth current gives $\psi>0$, so $\mu>0$ is an eigenvalue of the monodromy, and $2\log\mu$ is its signed logarithmic multiplier. When the monodromy is hyperbolic, its nonnegative rapidity is $2|\log\mu|$. This completes the proof.
\end{proof}

The zero mode of smooth near horizon currents enters the entropy formula of~\cite{Afshar:2016kjj}. For the particular singular Floquet current above, the principal value equals $\Delta$; extending the entropy interpretation to moving poles requires a separate symplectic and charge analysis. The poles themselves are dynamical, transported by the superrotations, and their number is the winding. What we have not done is to construct the symplectic structure on currents with moving poles. The charges $\oint Y\omega$ require the principal value prescription, and whether the resulting Poisson brackets acquire boundary terms at the poles, which would show up as anomalies in the charge algebra, is an open question. We are not aware of a treatment of such singular near horizon currents in the literature.

%======================================================================
\section{Twisted amplitudes}\label{sec:twisted}

Observables that resolve the cyclic label must project onto definite values of $r$. We construct the projectors, express the resulting amplitudes through orbital integrals, derive selection rules, and finally evaluate the amplitudes in closed form for Gaussian test functions, all on the global orbit where the measures are explicit.

\subsection{Projectors and the physical rotation}\label{sec:projectors}

To resolve the label we need a space that contains all its values at once. All values of $r$ occur together in the regular representation,
\[
L^2\big(\SL\big)=\bigoplus_{r\in\Z/2m\Z}\mathcal H^{\rm glob}_r ,
\]
on which the right translations $(R_h\psi)(g)=\psi(gh)$ commute with the left translations $(L_f\psi)(g)=\psi(f^{-1}g)$. The projectors are given by the following proposition.

\begin{proposition}\label{prop:projectors}
The operators $P_r=\frac1{2m}\sum_{\ell=0}^{2m-1}e^{i\pi r\ell/m}R_{\gamma^\ell}$ are orthogonal projectors commuting with the left action, with $\sum_rP_r=1$, $R_\gamma P_r=e^{-i\pi r/m}P_r$ and range $\mathcal H^{\rm glob}_r$. The central element $z=-I$ acts on $\mathcal H^{\rm glob}_r$ by $(-1)^r$ from both sides.
\end{proposition}

\begin{proof}
The identities follow from the orthogonality of the characters of $C_{2m}$ (appendix~\ref{app:tech}), and $P_r^*=P_r$ since $R_{\gamma^\ell}^*=R_{\gamma^{-\ell}}$. The range consists of the $\psi$ with $\psi(g\gamma)=e^{-i\pi r/m}\psi(g)$. Since $z$ is central, $L_z=R_z^{-1}$, and $R_z=R_\gamma^m$ acts by $(-1)^r$. This completes the proof.
\end{proof}

It is important to distinguish these projectors from those built from the physical rotation, which acts by left translation. The next proposition shows that the two notions differ.

\begin{proposition}\label{prop:leftright}
Let $m\ge2$. On $\mathcal H^{\rm glob}_r$ the left translation $L_\gamma$ is not a multiple of the identity. Its spectrum is $\{e^{i\pi s''/m}:s''\equiv r\pmod2\}$, which contains at least two values. Consequently the projector $\Pi_r=\frac1{2m}\sum_\ell e^{-i\pi r\ell/m}L_{\gamma^\ell}$ selects the states whose rest frame angular momentum lies in the class $r/2$ modulo $m$, and it does not select the monodromy sector.
\end{proposition}

\begin{proof}
By Theorem~\ref{thm:floquet}, $\mathcal H^{\rm glob}_r$ contains $\mathcal L_r=\Ind_K^{\SL}\sigma_r$, and by Proposition~\ref{prop:landau} it contains principal series representations of parity $r\bmod 2$. Each of them contains every $K$ type $s''\equiv r\pmod2$ for the left action of $K$, on which $L_\gamma=L_{k(\pi/m)}$ acts by $e^{i\pi s''/m}$. The values for $s''=r$ and $s''=r+2$ differ by $e^{2\pi i/m}\neq1$. Only weights of the parity of $r$ occur, since the central element acts by $(-1)^r$. The projector $\Pi_r$ is the spectral projector of the left rotation onto the weights $s''\equiv r\pmod{2m}$. This completes the proof.
\end{proof}

The distinction has a clear physical meaning. The left action of $\gamma$ is a rotation of the state through $2\pi/m$ in spacetime, and its eigenvalues are angular momentum phases, as in the twisted partition functions of orbifold theories. The right action of $\gamma$ is the action of the finite symmetry of the gravitational configuration on the internal state, and its eigenvalue is the monodromy of the line bundle. Only the central element $z$, the rotation through $2\pi$, acts in the same way on both sides, which is why the statistics of the states is tied to the parity of $r$. A measurement that resolves $r$ must couple to the configuration, not only to spacetime rotations. At the level of the full orbit the projectors have an equally natural meaning: inducing from the identity component of the stabilizer alone gives a representation that carries the right action of $C_{2m}$, and the $P_r$ decompose it into the sectors of definite $r$. Since left and right translations commute, the projectors $P_r$ and $\Pi_{r'}$ commute as well. Their joint ranges consist of states with a definite monodromy $r$ and a definite class $r'/2$ modulo $m$ of the rest frame angular momentum, and by Proposition~\ref{prop:leftright} several classes $r'$ occur for each $r$, subject only to $r'\equiv r\pmod2$.

\subsection{Orbital integrals and selection rules}\label{sec:orbital}

The branches are not admissible, so the operators of interest are not of trace class and their traces do not exist. We therefore work with finite quantities defined directly. For $\phi\in C_c^\infty(\SL)$ the operator $\pi_L(\phi)=\int\phi(g)L_g\,dg$ commutes with $R_h$, and by the unimodularity of $\SL$ the operator $R_h\pi_L(\phi)$ has kernel $\phi(xhy^{-1})$, whose diagonal is $\phi(xhx^{-1})$. For $h=\gamma^\ell$ with $\ell\notin\{0,m\}$ the integral of this diagonal is finite, as the following theorem shows.

\begin{theorem}\label{thm:orbital}
For $\ell\notin\{0,m\}$, $\gamma^\ell=k(\pi\ell/m)$ is regular elliptic with centralizer $K$, and
\begin{equation}\label{eq:Wl}
W_\ell(\phi)=\int_{\SL}\phi\big(x\gamma^\ell x^{-1}\big)dx=\mathrm{vol}(K)\int_{\SL/K}\phi(x\gamma^\ell x^{-1})\,d\dot x
\end{equation}
is a finite elliptic orbital integral for every $\phi\in C_c^\infty(\SL)$.
\end{theorem}

\begin{proof}
A matrix commuting with $k(\vartheta)$, $\vartheta\notin\pi\Z$, preserves its complex eigenlines and is a rotation, so the centralizer is $K$, and the integral factorizes over $\SL/K$. The conjugacy class of $\gamma^\ell$ is a component of $\{\tr g=2\cos(\pi\ell/m)\}$, the components being distinguished by the sign of the lower left entry, which cannot vanish when $|\tr g|<2$. It is therefore closed. A closed orbit of a smooth action of a Lie group is an embedded submanifold, and the orbit map from the quotient by the stabilizer is a diffeomorphism onto it. Hence $xK\mapsto x\gamma^\ell x^{-1}$ is a homeomorphism of $\SL/K$ onto a closed subset of $\SL$, and in particular proper, so the preimage of the support of a compactly supported $\phi$ is compact. With the normalization of section~\ref{sec:explicit}, $dx=d\mu(\dot x)\,dk$ with $\mathrm{vol}(K)=2\pi$, and since the integrand is invariant under $x\mapsto xk$, the integral over $\SL$ equals $\mathrm{vol}(K)$ times the integral of a compactly supported continuous function over $\SL/K$, which is finite. This completes the proof.
\end{proof}

For $\ell\in\{0,m\}$ the diagonal is the constant $\phi(\pm I)$, whose integral over $\SL$ diverges unless $\phi(\pm I)=0$. We never assign a value to such an integral. Instead we define the finite monodromy resolved amplitude
\begin{equation}\label{eq:Zr}
Z^\flat_r(\phi)=\frac1{2m}\sum_{\ell\notin\{0,m\}}e^{i\pi r\ell/m}W_\ell(\phi),
\end{equation}
where $\ell$ runs over $\Z/2m\Z$. Formally, the diagonal of the kernel of $P_r\pi_L(\phi)$ integrates to $Z^\flat_r(\phi)$ plus a term proportional to $\phi(I)+(-1)^r\phi(-I)$ with an infinite coefficient. This formal term vanishes when $\phi(\pm I)=0$, and it is the same for $r$ and $r'$ when $r\equiv r'\pmod2$. The meaningful quantities are therefore
\begin{equation}\label{eq:Zrsplit}
Z_r(\phi)=Z^\flat_r(\phi)\quad\text{if }\phi(\pm I)=0,\qquad \Delta(r,r';\phi)=Z^\flat_r(\phi)-Z^\flat_{r'}(\phi)\quad\text{if }r\equiv r'\pmod2 .
\end{equation}
The second quantity is a single finite functional of $\phi$, and no individual $Z_r(\phi)$ is defined when $\phi(\pm I)\neq0$. We call $W_\ell$, $Z_r$ and $\Delta$ finite orbital amplitudes: $\phi$ is a test function, and we do not claim a trace identity for any operator.

The whole dependence of these amplitudes on $r$ beyond parity is carried by elliptic orbital integrals at the angles $\pi\ell/m$. In spacetime $\gamma^\ell$ rotates through $\vartheta=2\pi\ell/m$ about a timelike axis, with transverse fixed point factor $(4\sin^2(\vartheta/2))^{-1}$, the factor of the one loop computation~\cite{Barnich:2015mui}. The amplitudes have a structural resemblance to orbifold twisted terms, with the twist acting on the right. They are finite monodromy resolved orbital functionals on the global branch, not trace characters or gravitational partition functions.

\label{sec:selection}Besides the finite part~\eqref{eq:Zrsplit}, the cyclic structure imposes arithmetic constraints that do not depend on the regulator. The first comes from the centre.

\begin{proposition}\label{prop:parity}
Let $v$ lie in a subrepresentation of the left regular representation of $\SL$ on which $z$ acts by $(-1)^{2j}$. Then $P_rv=0$ unless $r\equiv2j\pmod2$.
\end{proposition}

\begin{proof}
Since $P_r$ commutes with $L_z$, $(-1)^{2j}P_rv=L_zP_rv=(-1)^rP_rv$. This completes the proof.
\end{proof}

A stronger rule holds for amplitudes obtained from a kernel built by a method of images, which is the typical situation in one loop computations. Suppose that such a construction gives finite numbers $W_\ell=\sum_{n\ge1}w_n(\gamma^{n\ell})$ for all $\ell\in\Z/2m\Z$, including $\ell\in\{0,m\}$, with class functions $w_n$, and set $Z_r=\frac1{2m}\sum_{\ell\in\Z/2m\Z}e^{i\pi r\ell/m}W_\ell$ for these amplitudes. Here $n$ labels images and should not be confused with the winding of loops in section~\ref{sec:topology}.

\begin{proposition}\label{prop:images}
Let $d_n=\gcd(n,2m)$. The $n$th image contributes to $Z_r$ only if $d_n\mid r$. If moreover $w_n(gz^n)=(-1)^{2jn}w_n(g)$, it contributes only if $r\equiv2jn\pmod2$.
\end{proposition}

\begin{proof}
The function $\ell\mapsto w_n(\gamma^{n\ell})$ is constant on the cosets of the kernel of the map $\ell\mapsto n\ell$ on $\Z/2m\Z$. This kernel consists of the $d_n$ elements $\ell=2mk/d_n$ with $0\le k<d_n$, and summing over it gives the factor
\[
\sum_{k=0}^{d_n-1}e^{2\pi irk/d_n},
\]
which vanishes unless $d_n\mid r$. Pairing $\ell$ with $\ell+m$ multiplies the phase by $(-1)^r$ and the summand by $(-1)^{2jn}$, so the pair cancels unless the signs agree. This completes the proof.
\end{proof}

For $m=3$, consider an illustrative method of images built from a spin one half elementary factor. We make the additional assumption that the $n$th image term acquires the central sign $(-1)^n$ (equivalently $2j_n\equiv n\pmod 2$), rather than imposing spin one half independently on every image term. Proposition~\ref{prop:images} then gives the following residues modulo $6$:
\[
\begin{array}{c|cccccc}
 n &1&2&3&4&5&6\\\hline
 d_n=\gcd(n,6)&1&2&3&2&1&6\\
 2j_n\bmod2&1&0&1&0&1&0\\
 r\bmod6&\{1,3,5\}&\{0,2,4\}&\{3\}&\{0,2,4\}&\{1,3,5\}&\{0\}
\end{array}
\]
The table is periodic in $n$ modulo $6$. The assumed central sign is an additional property of the selected image construction, not a consequence of the cyclic little group alone. If instead one fixes odd central parity separately for every term, the even $n$ columns vanish, as required by Proposition~\ref{prop:images}. The divisibility and parity restrictions have different origins: the first comes from the kernel of multiplication by $n$, the second from the central action in the spin cover. For $m=3$ the single image term of the one loop result $\log Z=\sum_k\frac1k2\,\mathrm{Re}\frac{q^{2k}}{1-q^k}$~\cite{Barnich:2015mui}, namely $2\,\mathrm{Re}\frac{q^2}{1-q}=-(1+2\cos\vartheta)$ with $q=e^{i\vartheta}$, vanishes at $\vartheta=2\pi/3,4\pi/3$. The local fixed axis term thus does not distinguish the nontrivial elements of $C_6$, and differences between these image amplitudes $Z_r$ must come from nonlocal parts of the orbital integrals. Rotations of finite order have fixed lines, and the vacuum character diverges at rational angles, so a regulator such as $\vartheta\to\vartheta+i\epsilon$ is implicit in such comparisons.

\subsection{Explicit amplitudes}\label{sec:explicit}

The selection rules constrain the amplitudes without computing them. We now show that for test functions built from the norm of a matrix the orbital integrals and the finite amplitudes $Z_r$ and $\Delta$ of~\eqref{eq:Zrsplit} can be evaluated in closed form, that their dependence on $r$ is universal within this class, and that they resolve the cyclic label. Write $a(t)=\mathrm{diag}(e^{t/2},e^{-t/2})$, so that every $g\in\SL$ has the form $k(\alpha)a(t)k(\psi)$ with $t\ge0$, and normalize the Haar measure by
\[
\int_{\SL}f(g)\,dg=\frac12\int_0^{2\pi}\!\!\int_0^\infty\!\!\int_0^{2\pi}f\big(k(\alpha)a(t)k(\psi)\big)\sinh t\,d\alpha\,dt\,d\psi ,
\]
where the factor $\frac12$ compensates for $k(\alpha)a(t)k(\psi)=k(\alpha+\pi)a(t)k(\psi+\pi)$. In this normalization $\mathrm{vol}(K)=2\pi$, and the invariant measure on $\Hb^2=\SL/K$ is $d\mu=\frac12\sinh t\,d\alpha\,dt$ with $0\le\alpha<2\pi$, consistently with Theorem~\ref{thm:orbital} and with $dg=d\mu(\xi)\,d\omega$ in section~\ref{sec:shift}. For a function of $gK$ the $\psi$ integral gives $\mathrm{vol}(K)=2\pi$, which is the relation $dg=d\mu\,d\omega$ with $0\le\omega<2\pi$. The same normalization is used throughout the paper, and the coefficient $2\pi^2=\frac12\cdot2\pi\cdot2\pi$ in the proof of Proposition~\ref{prop:gaussW} comes from the integrations over $\psi$ and $\alpha$. Let $\|g\|^2=\tr(g^Tg)$, so that $\|k(\alpha)a(t)k(\psi)\|^2=2\cosh t\ge2$, and let $J=k(\pi/2)$. For a measurable profile $F$ on $[2,\infty)$ with $\int_2^\infty|F(x)|\,(1+x)\,dx<\infty$ we consider
\[
\phi_0^F(g)=F\big(\|g\|^2\big),\qquad\phi_1^F(g)=\tr(Jg)\,F\big(\|g\|^2\big),\qquad\phi_2^F(g)=\tr(g)\tr(Jg)\,F\big(\|g\|^2\big),
\]
and we set $I_F=\int_0^\infty F(2+v)\,dv$. The Gaussian profile $F(x)=e^{-\beta x}$ with $\beta>0$ gives $I_F=e^{-2\beta}/\beta$.

Haar measures are fixed only up to a constant, so the normalization is a convention, but it must be used consistently. The following lemma identifies it.

\begin{lemma}\label{lem:haar}
In the coordinates $g=k(\alpha)a(t)k(\psi)$ the left invariant volume form $|\omega_{11}\wedge\omega_{12}\wedge\omega_{21}|$ built from $g^{-1}dg=(\omega_{ij})$ equals $\sinh t\,d\alpha\,dt\,d\psi$. The map $(\alpha,t,\psi)\mapsto g$ from $[0,2\pi)\times(0,\infty)\times[0,2\pi)$ is two to one onto the set $\{t>0\}$, whose complement $K$ has measure zero.
\end{lemma}

\begin{proof}
Let $E=\begin{pmatrix}0&1\\-1&0\end{pmatrix}$ and $H=\frac12\mathrm{diag}(1,-1)$, so that $k(\alpha)^{-1}dk(\alpha)=E\,d\alpha$ and $a(t)^{-1}da(t)=H\,dt$. Since $E$ commutes with $k(\psi)$,
\[
g^{-1}dg=k(\psi)^{-1}\Big(a(t)^{-1}Ea(t)\,d\alpha+H\,dt+E\,d\psi\Big)k(\psi),\qquad a(t)^{-1}Ea(t)=\begin{pmatrix}0&e^{-t}\\-e^{t}&0\end{pmatrix}.
\]
Conjugation by $k(\psi)$ is a linear map of $\sltwo$ with determinant one, because $\SL$ is connected and its adjoint representation preserves the Killing form, so it does not change the volume form. In the coordinates $(X_{11},X_{12},X_{21})$ of a traceless matrix the three coefficients are $(0,e^{-t},-e^{t})$, $(\frac12,0,0)$ and $(0,1,-1)$, and the determinant of these three rows is $-\frac12(e^{t}-e^{-t})=-\sinh t$. Hence the volume form is $\sinh t\,d\alpha\,dt\,d\psi$. For $t>0$ the decomposition $g=k(\alpha)a(t)k(\psi)$ is a singular value decomposition with distinct singular values $e^{\pm t/2}$, which determine $t$. The singular vectors then determine $k(\alpha)$ and $k(\psi)$ up to a simultaneous change of sign of corresponding columns, and since both factors lie in $SO(2)$ the only possibility is the replacement of both by their negatives, that is $(\alpha,\psi)\mapsto(\alpha+\pi,\psi+\pi)$. This gives exactly two preimages. This completes the proof.
\end{proof}

Hence $\int_{\SL}f\,dg=\frac12\int\!\!\int\!\!\int f\sinh t\,d\alpha\,dt\,d\psi$ is the Haar measure of the Maurer and Cartan volume form.

The normalization and the prefactor $2\pi^2$ can be derived a second time in Iwasawa coordinates, which do not involve the double cover. Write $g=n(x)a(s)k(\theta)$ with $n(x)=\begin{pmatrix}1&x\\0&1\end{pmatrix}$, $x,s\in\R$ and $0\le\theta<2\pi$, a parametrization that is one to one. Since $n(x)^{-1}dn(x)=E_{12}\,dx$ and $a(s)^{-1}E_{12}a(s)=e^{-s}E_{12}$, the same computation as in the proof of Lemma~\ref{lem:haar} gives the coefficient rows $(0,e^{-s},0)$, $(\frac12,0,0)$ and $(0,1,-1)$, with determinant $\frac12e^{-s}$. Hence $dg=\frac12e^{-s}\,dx\,ds\,d\theta$ for the same Maurer and Cartan form, and
\[
W_\vartheta(\phi)=\pi\int_{\R}\int_{\R}\phi\big(y\,k(\vartheta)\,y^{-1}\big)\,e^{-s}\,dx\,ds,\qquad y=n(x)a(s),
\]
because the integrand does not depend on $\theta$. This expression contains neither the factor $\frac12$ of the double cover nor the factor $2\pi^2$, and it gives an independent route to Proposition~\ref{prop:gaussW} below.

The functions $\phi_i^F$ are not compactly supported. Since $|\tr g|$ and $|\tr(Jg)|$ are bounded by $\sqrt2\|g\|$, one has $|\phi_i^F(g)|\le2(1+x)|F(x)|$ with $x=\|g\|^2$, and $dx=2\sinh t\,dt$ shows that the $\phi_i^F$ lie in $L^1(\SL)$. Thus $\pi_L(\phi_i^F)$ is bounded by Young's inequality, and $R_h\pi_L(\phi_i^F)$ has the measurable convolution kernel $\phi_i^F(xhy^{-1})$, in the sense that $(R_h\pi_L(\phi_i^F)f)(x)=\int\phi_i^F(xhy^{-1})f(y)\,dy$ for $f\in L^2(\SL)$. This identity holds for almost every $x$, where the integral converges absolutely, and both sides define the same bounded operator on $L^2(\SL)$ by Young's inequality. An integrable test function does not give a trace class operator, and no trace identity is used. The absolute convergence of the orbital integrals does not follow from integrability on $\SL$ and is established separately from the explicit formulas of Proposition~\ref{prop:gaussW}. These integrals converge absolutely, so~\eqref{eq:Wl},~\eqref{eq:Zr} and~\eqref{eq:Zrsplit} apply to them. Since these functions are not in $C_c^\infty(\SL)$, Theorem~\ref{thm:orbital} does not apply as stated, and the finiteness of $W_\ell$ for $\ell\neq0,m$ follows instead from the explicit formulas below. The only regularity used is that $F$ is measurable with $\int_2^\infty|F(x)|(1+x)\,dx<\infty$. Under $g\mapsto-g$, $\phi_0^F$ and $\phi_2^F$ are even and $\phi_1^F$ is odd. Under the reflection $g\mapsto\epsilon g\epsilon$ with $\epsilon=\mathrm{diag}(1,-1)$, which maps $k(\vartheta)$ to $k(-\vartheta)$ and preserves the Haar measure, $\phi_0^F$ is even while $\phi_1^F$ and $\phi_2^F$ are odd, because $\epsilon J\epsilon=-J$. Thus $\phi_1^F$ and $\phi_2^F$ detect the sense of rotation. The orbital integrals are given by the following proposition.

\begin{proposition}\label{prop:gaussW}
For $\vartheta\notin\pi\Z$ let $W_\vartheta(\phi)=\int_{\SL}\phi(xk(\vartheta)x^{-1})\,dx$ and $\sigma=\sin\vartheta$. Then
\[
W_\vartheta(\phi_0^F)=\frac{\pi^2}{2|\sigma|}\int_0^\infty\frac{F(2+v)}{\sqrt{4\sigma^2+v}}\,dv,\qquad
W_\vartheta(\phi_1^F)=-\frac{\pi^2I_F}{2\sin\vartheta},\qquad
W_\vartheta(\phi_2^F)=-\pi^2I_F\cot\vartheta .
\]
For the Gaussian profile $W_\vartheta(\phi_0^F)=\frac{\pi^{5/2}}{2\sqrt\beta\,|\sigma|}\,e^{-2\beta\cos2\vartheta}\,\mathrm{erfc}\big(2\sqrt\beta\,|\sigma|\big)$.
\end{proposition}

\begin{proof}
All integrands are measurable, and we first carry out the computation below with $|F|$ in place of $F$ and absolute values of the trace factors. The resulting iterated integrals are bounded by $\frac{\pi^2}{4\sigma^2}\int_0^\infty|F(2+v)|\,dv$ for $\phi_0^F$, by $\frac{\pi^2}{2|\sigma|}\int_0^\infty|F(2+v)|\,dv$ for $\phi_1^F$ and by twice this for $\phi_2^F$, which are finite under our assumption on $F$. By Tonelli's theorem, $x\mapsto\phi_i^F(xk(\vartheta)x^{-1})$ is therefore absolutely integrable over $\SL$, and Fubini's theorem justifies the same steps for $\phi_i^F$. Only $\int_0^\infty|F(2+v)|\,dv<\infty$ is needed here. The stronger weight $1+x$ in our assumption is used only for the integrability of $\phi_i^F$ on $\SL$. Write $x=k(\alpha)a(t)k(\psi)$. Since $k(\psi)$ commutes with $k(\vartheta)$, and conjugation by $k(\alpha)$ preserves $\|\cdot\|$, $\tr$ and $\tr(J\,\cdot)$ because $J\in K$, the integrand depends only on $g_t=a(t)k(\vartheta)a(t)^{-1}$, whose entries are $\cos\vartheta$, $e^t\sin\vartheta$, $-e^{-t}\sin\vartheta$, $\cos\vartheta$. Hence
\[
\|g_t\|^2=2+4\sigma^2\sinh^2t,\qquad\tr(g_t)=2\cos\vartheta,\qquad\tr(Jg_t)=-2\sigma\cosh t ,
\]
and $W_\vartheta(\phi)=2\pi^2\int_0^\infty\phi(g_t)\sinh t\,dt$. We substitute
\[
v=4\sigma^2\sinh^2t,\qquad dv=8\sigma^2\sinh t\cosh t\,dt,\qquad\cosh t=\sqrt{1+v/4\sigma^2}.
\]
For $\phi_1^F$ the factor $\cosh t$ cancels and
\[
W_\vartheta(\phi_1^F)=2\pi^2\int_0^\infty(-2\sigma\cosh t)F(2+v)\,\frac{dv}{8\sigma^2\cosh t}=-\frac{\pi^2}{2\sigma}\,I_F .
\]
The function $\phi_2^F$ carries the additional constant factor $2\cos\vartheta$. For $\phi_0^F$ the same substitution gives $\frac{\pi^2}{4\sigma^2}\int_0^\infty F(2+v)(1+v/4\sigma^2)^{-1/2}dv$, which is the stated formula. All integrals converge absolutely, because $\int_0^\infty|F(2+v)|\,dv<\infty$ and $(1+v/4\sigma^2)^{-1/2}\le1$. For $F(x)=e^{-\beta x}$, the identity $\int_0^\infty e^{-\beta v}(4\sigma^2+v)^{-1/2}dv=e^{4\beta\sigma^2}\sqrt{\pi/\beta}\,\mathrm{erfc}(2\sqrt\beta|\sigma|)$ and $-2\beta+4\beta\sigma^2=-2\beta\cos2\vartheta$ give the last formula. This completes the proof.
\end{proof}

The factor $1/\sin\vartheta$ is the Weyl denominator that appears in the characters of the discrete series on the elliptic set~\cite{Knapp:1986}. The striking feature is that for $\phi_1^F$ and $\phi_2^F$ the profile enters only through the constant $I_F$. Inserting the proposition into~\eqref{eq:Zr} with $\vartheta=\pi\ell/m$ gives the amplitudes in closed form, as stated in the following theorem.

\begin{theorem}\label{thm:gaussZ}
Let $m\ge2$ and represent $r$ by an integer $0\le r\le2m-1$. Then $\phi_1^F(\pm I)=\phi_2^F(\pm I)=0$, and the amplitudes $Z_r(\phi_1^F)$ and $Z_r(\phi_2^F)$ are finite. For odd $r$,
\[
Z_r(\phi_1^F)=-\frac{i\pi^2I_F}{2m}\,(m-r),\qquad Z_r(\phi_2^F)=0 .
\]
For even $r$ with $2\le r\le2m-2$,
\[
Z_r(\phi_1^F)=0,\qquad Z_r(\phi_2^F)=-\frac{i\pi^2I_F}{m}\,(m-r),
\]
and $Z_0(\phi_1^F)=Z_0(\phi_2^F)=0$. For $r\equiv r'\pmod2$ the finite functional $\Delta(r,r';\phi_0^F)$ of~\eqref{eq:Zrsplit} vanishes for odd $r,r'$, and for even $r,r'$ it equals
\[
\Delta(r,r';\phi_0^F)=\frac1m\sum_{\ell=1}^{m-1}\Big(\cos\frac{\pi r\ell}m-\cos\frac{\pi r'\ell}m\Big)W_{\pi\ell/m}(\phi_0^F).
\]
\end{theorem}

\begin{proof}
Since $\tr(\pm J)=0$, $\phi_1^F$ and $\phi_2^F$ vanish at $\pm I$, so $Z_r=Z^\flat_r$ by~\eqref{eq:Zrsplit}. For $\phi=\phi_1^F$ we use $\gamma^{\ell+m}=-\gamma^\ell$ and the oddness of $\phi_1^F$ to pair $\ell$ with $\ell+m$, which gives
\[
Z_r(\phi_1^F)=\frac{1-(-1)^r}{2m}\sum_{\ell=1}^{m-1}e^{i\pi r\ell/m}\,W_{\pi\ell/m}(\phi_1^F).
\]
This vanishes for even $r$. For odd $r$ the substitution $\ell\mapsto m-\ell$ leaves $W_{\pi\ell/m}(\phi_1^F)$ unchanged and replaces the phase by $-e^{-i\pi r\ell/m}$, so the sum equals $i\sum_{\ell=1}^{m-1}\sin\frac{\pi r\ell}m\,W_{\pi\ell/m}(\phi_1^F)$. For odd $r=2p+1\le2m-1$ one has $\frac{\sin rx}{\sin x}=1+2\sum_{j=1}^p\cos2jx$, and $\sum_{\ell=1}^{m-1}\cos\frac{2\pi j\ell}m=-1$ for $1\le j\le p\le m-1$, hence
\[
\sum_{\ell=1}^{m-1}\frac{\sin(\pi r\ell/m)}{\sin(\pi\ell/m)}=(m-1)-2p=m-r ,
\]
and Proposition~\ref{prop:gaussW} gives the first formula. For $\phi=\phi_2^F$, which is even, the same pairing gives the factor $1+(-1)^r$, so $Z_r(\phi_2^F)=0$ for odd $r$. For even $r$ the substitution $\ell\mapsto m-\ell$ changes the sign of $\cot(\pi\ell/m)$ and replaces the phase by $e^{-i\pi r\ell/m}$, and the sum becomes $i\sum_\ell\sin\frac{\pi r\ell}m\cot\frac{\pi\ell}m$. It vanishes for $r=0$. For $2\le r\le2m-2$, writing $2\sin rx\cos x=\sin(r+1)x+\sin(r-1)x$ and applying the previous identity to the odd integers $r\pm1\in[1,2m-1]$ gives $\frac12\big((m-r-1)+(m-r+1)\big)=m-r$, which is the second formula. For $\phi_0^F$ the functional is $\Delta(r,r';\phi_0^F)=Z^\flat_r-Z^\flat_{r'}$ by~\eqref{eq:Zrsplit}. Since $\phi_0^F$ is even and $W_{\pi\ell/m}(\phi_0^F)$ depends only on $|\sin(\pi\ell/m)|$, the pairing gives the factor $1+(-1)^r$, and the substitution $\ell\mapsto m-\ell$ turns the phases into cosines for even $r$. This completes the proof.
\end{proof}

The theorem and the preceding proposition are proved for all admissible profiles. The numerical tests described in appendix~\ref{app:tech}, for three profiles and finitely many $m$ and angles, support these formulas but do not prove them. The theorem gives the explicit dependence on $r$, and for the orientation odd test functions this dependence is universal: every profile gives the same function $m-r$, up to the factor $I_F$. The amplitudes of $\phi_1^F$ and $\phi_2^F$ are odd under $r\mapsto-r$, as required by their oddness under the reflection, and they are linear in the representative $r$, which is reminiscent of the linear dependence of the formal degrees of the discrete series on the lowest weight. These properties give a complete separation of the cyclic sectors, stated in the following corollary. For a label $r$ let $r_0\in\{0,1\}$ be its parity, so that $\Delta(r,r_0;\phi)$ is defined.

\begin{corollary}\label{cor:separation}
Let $m\ge2$, and let $F\ge0$ be measurable with $\int_2^\infty(1+x)F(x)\,dx<\infty$ and $F\neq0$ on a set of positive measure. Then the map
\[
r\longmapsto\Big(r_0,\ Z_r(\phi_1^F),\ Z_r(\phi_2^F),\ \Delta(r,r_0;\phi_0^F)\Big)
\]
is injective on $\Z/2m\Z$.
\end{corollary}

\begin{proof}
Since $F\ge0$ is not almost everywhere zero, $I_F>0$, and $W_\vartheta(\phi_0^F)>0$ for every $\vartheta\notin\pi\Z$ by Proposition~\ref{prop:gaussW}. Let $r\neq r'$ have the same image. Then $r_0=r'_0$. If both are odd, Theorem~\ref{thm:gaussZ} gives $I_F(m-r)=I_F(m-r')$ with $r,r'\in\{1,\dots,2m-1\}$, so $r=r'$, a contradiction. If both are even, $Z_r(\phi_2^F)=Z_{r'}(\phi_2^F)$ and Theorem~\ref{thm:gaussZ} show that $m-r=m-r'$ when $r,r'\neq0$, and that the other label is $m$ when one of them is $0$, which requires $m$ to be even. It remains to separate $r=0$ from $r=m$ for even $m$. Here $\Delta(0,0;\phi_0^F)=0$, while by Theorem~\ref{thm:gaussZ}
\[
\Delta(m,0;\phi_0^F)=\frac1m\sum_{\ell=1}^{m-1}\big((-1)^\ell-1\big)W_{\pi\ell/m}(\phi_0^F)=-\frac2m\sum_{\ell\ \rm odd}W_{\pi\ell/m}(\phi_0^F)<0 .
\]
For odd $m$ the labels $0$ and $m$ have different parity and are separated by $r_0$. This completes the proof.
\end{proof}

In particular the Gaussian profile separates all $2m$ sectors. Only the pair $\{0,m\}$ for even $m$ needs the even profile $\phi_0^F$, because both labels are fixed by $r\mapsto-r$, so the orientation odd amplitudes vanish on them. Three consistency checks are immediate. The vanishing of $Z_r(\phi_1^F)$ for even $r$ is Proposition~\ref{prop:parity} for a function odd under the centre. Moreover $\sum_rZ^\flat_r(\phi)=0$ for every $\phi$, because $\sum_{r\in\Z/2m\Z}e^{i\pi r\ell/m}=0$ for $\ell\neq0$, and indeed $\sum_{r\ \rm odd}(m-r)=0$ and $\sum_{r\ \rm even,\,r\neq0}(m-r)=0$. Finally, for real $F$, $\overline{Z_r}=Z_{-r}$, which together with $Z_{-r}=-Z_r$ explains why these amplitudes are purely imaginary. For $m=3$, for example, $Z_r(\phi_1^F)$ is proportional to $2,0,-2$ for $r=1,3,5$, and $Z_r(\phi_2^F)$ is proportional to $0,1,-1$ for $r=0,2,4$.

Functions of $\|g\|^2$ are functions of the radial coordinate $t$ alone, that is, invariant under left and right multiplication by $K$. They were chosen because their orbital integrals are elementary while keeping the essential features: the centre acts by a definite sign, and $\tr(Jg)$ is the simplest factor that detects orientation. The heat kernel of a left invariant Riemannian metric on $\SL$ is invariant only under conjugation by $K$, not under $K\times K$, and its orbital integrals at elliptic elements involve a genuine two variable integral, which we have not evaluated. By the universality just found we expect the dependence on $r$ of its orientation odd part to be again proportional to $m-r$, but we have not proved this. A comparison with the Wilson spool~\cite{Pannier:2025wil}, which organizes one loop determinants as sums over windings of Wilson loops, would require the full heat kernel and is left for future work.

%======================================================================
\section{Orbit topology}\label{sec:topology}

Under the smooth principal bundle hypothesis below, a cyclic character that is trivial on the connected stabilizer can be interpreted as the holonomy of an associated flat line bundle. The required smooth quotient topology of the full orbit has not been established here. We therefore state the result under the assumption of section~\ref{sec:pullback}, in the form that $\widetilde G\to\Om_{m,\Delta}$ is a locally trivial principal bundle with fibre $\widetilde H$. Its total space $\widetilde G$ is contractible, which leads to the following result.

\begin{proposition}\label{prop:pi1}
Assume that $\widetilde G\to\Om_{m,\Delta}$ is a locally trivial principal bundle with fibre $\widetilde H$. Then $\pi_1(\Om_{m,\Delta})\cong\pi_0(\widetilde H)\cong\Z$. For $q$ invariant under $R_m$ a generator is the loop obtained by rotating $q$ rigidly through $2\pi/m$, and the full rotation represents $m$ times the generator.
\end{proposition}

\begin{proof}
A locally trivial bundle is a Serre fibration, so its homotopy sequence is exact, and it gives $\pi_1(\Om_{m,\Delta})\cong\pi_0(\widetilde H)$ since $\widetilde G$ is contractible, and $\pi_0(\widetilde H)\cong\Z$ by~\eqref{eq:Htilde}, generated by $\widetilde R_m$. The connecting map sends the path of translations from the identity to $\widetilde R_m$ to the rotation of $q$ through $2\pi/m$, and $T=\widetilde R_m^{\,m}$. This completes the proof.
\end{proof}

Under the same assumption, the homotopy sequence also gives
\[
\pi_1(G)\longrightarrow\pi_1(\Om_{m,\Delta})\longrightarrow\pi_0(H)\longrightarrow0.
\]
Here the full rotation maps to $m$ and $\pi_0(H)\cong\Z_m$ is the cokernel. A symmetric configuration can thus be carried around a noncontractible loop by a rotation through $2\pi/m$ that leaves it unchanged at the end. Flat line bundles are classified by $\mathrm{Hom}(\Z,U(1))\cong U(1)$. We now make the construction and the conventions explicit. For a character $\chi$ of $\widetilde H$ the homogeneous line bundle is $L_\chi=\widetilde G\times_{\widetilde H}\C_\chi$, the quotient of $\widetilde G\times\C$ by $(gh,z)\sim(g,\chi(h)z)$. If $\chi$ is trivial on the identity component $\widetilde H^0$, then $L_\chi$ is associated with the covering $\widetilde G/\widetilde H^0\to\widetilde G/\widetilde H=\Om_{m,\Delta}$, whose deck group $\pi_0(\widetilde H)\cong\Z$ acts by right multiplication. The local sections of $L_\chi$ that lift to locally constant functions on $\widetilde G/\widetilde H^0$ define a flat connection, which is invariant under $\widetilde G$ because the left action commutes with the deck group. We orient the loop defined by $h\in\pi_0(\widetilde H)$ as the image of a path from $g\widetilde H^0$ to $gh\widetilde H^0$, and with this orientation its holonomy is $\chi(h)$, as in appendix~\ref{app:tech}. For $\chi=\rho_{0,r}$ and its universal cover version with $\chi(\widetilde R_m)=e^{i\alpha}$, the holonomy along the generator of Proposition~\ref{prop:pi1} is $e^{i\alpha}$. For $s\neq0$ two notions of equivalence must be distinguished. As homogeneous bundles, $L_\chi$ and $L_{\chi'}$ are isomorphic if and only if $\chi=\chi'$, since the isotropy representation on the fibre over the base point is an invariant, so different values of $s$ give inequivalent homogeneous bundles. As topological line bundles they are isomorphic, provided $\Om_{m,\Delta}$ is in addition paracompact, because the characters with parameter $ts$, $0\le t\le1$, define a line bundle over $\Om_{m,\Delta}\times[0,1]$, and line bundles over a paracompact base are homotopy invariant. An invariant connection on $L_\chi$ is determined by a linear form $\Lambda$ on the Lie algebra $\Vect$ that equals $d\chi$ on the Lie algebra of $\widetilde H$ and is invariant under $\widetilde H$, and its curvature at the base point is $-\Lambda([X,Y])$. This is Wang's description of invariant connections~\cite{Wang:1958}, whose extension to the present Fr\'echet setting we assume. Since $\Vect$ is perfect, flatness forces $\Lambda=0$ and hence $d\chi=0$, that is $s=0$. Thus for $s\neq0$ no invariant connection is flat. A bundle descends to $G^{(N)}$ if and only if $e^{imN\alpha}=1$, which on the double cover gives $\alpha=\pi r/m$. Equivariance, not flatness, quantizes the holonomy, and a loop of class $k$ is weighted by
\begin{equation}\label{eq:topweight}
e^{i\pi rk/m}.
\end{equation}
Under the smooth principal bundle hypothesis and for $s=0$, the cyclic phase $r$ is Fourier dual to the winding of loops and is not itself a winding number. For $s\ne0$ no invariant flat connection has been constructed by this argument. In a path integral over histories in $\Om_{m,\Delta}$~\cite{Merbis:2019wgk,Maxfield:2026men}, the flat factor adds the topological term~\eqref{eq:topweight}, and formally $\mathcal Z^{(r)}=\sum_ke^{i\pi rk/m}\mathcal Z_k$ with $\mathcal Z_k$ the contribution of winding class $k$. This formal partition function is unrelated to the finite orbital amplitudes $Z_r$ of section~\ref{sec:twisted}, and no gravitational path integral identity or superselection sector is constructed here. It is formal in this sense, which is also the precise sense of the analogy with a $\theta$ angle. Maxfield's exact path integrals over $T_{n,\Delta}$~\cite{Maxfield:2026men} use the one parameter subgroup as gauge group and find a regulator dependent overall constant with a possible fixed phase. Integrating with the full stabilizer instead would formally produce this sum over winding classes, and the cyclic label would offer a natural parametrization of the phase. We have not carried out this computation. The picture is not in conflict with the string like carriers of~\cite{Ruzziconi:2026ext}, and we do not identify $r$ with a target space winding number.

%======================================================================
\section{Discussion}\label{sec:discussion}

At cyclically symmetric points of hyperbolic winding Virasoro orbits, the formal smooth BMS$_3$ inducing construction restricts to the same finite shift condition as the finite cyclic sectors of $B(2,1)$. An independent unitary global branch exists, but its embedding into a full unitary BMS$_3$ representation has not been constructed. The mechanism is the direct product structure of their stabilizer, which makes the monodromy $r$ an independent label of the inducing data, whereas on the orbits of massive particles the same cyclic group lies in the circle of rotations and its label is fixed by the spin.

The physical picture that emerges is the following. The winding cosmology-type Bondi data yield locally flat geometries whose Lorentz holonomy is hyperbolic, as for a flat space cosmology, but whose boundary projective structure has nonzero winding. Their global completion and causal structure remain unanalysed. Its asymptotic symmetries include a hyperbolic flow and a finite group of rigid rotations, and the inducing data built on it carry, besides the continuous label of the flow, a phase under the rotations. The phase is additional inducing data and is unaffected by the splitting of the central extension considered in section~\ref{sec:chars}. Subject to the principal bundle assumption of section~\ref{sec:topology}, for a character trivial on the connected stabilizer it is the holonomy of a flat line bundle, Fourier dual to loop winding. This does not assert an invariant flat connection for nonzero continuous character $s$. Seen from the Poincar\'e group, at invariant points with timelike projected momentum, it fixes the offset of an infinite ladder of helicities, while the classical symmetry fixes its spacing. The Poincar\'e Casimir is constant on each global Lorentz orbit, and only the projected rest energy varies between different symmetric points of the same Virasoro orbit. The configurations of the orbit have unbounded negative energy, so the corresponding representations do not describe stable particles. These statements concern inducing data and the unitary global branches. A full Hilbert space realization of the representation of the infinite dimensional orbit is not constructed here. Whether these inducing labels give distinct physical sectors of a completed quantum gravity theory remains an open problem. The analogy with a topological angle is conditional and no gravitational path integral over the full orbit is computed. Several questions remain open.

Two of the questions left open by the general analysis have been settled here. The stabilizers of the parabolic orbits with winding have the same direct product structure as the hyperbolic ones (Proposition~\ref{prop:parabolic}), so the cyclic label is an independent inducing label on them as well, and for test functions of the matrix norm the monodromy resolved amplitudes are elementary functions of $r$, universal for the orientation odd ones, that distinguish all sectors (Theorem~\ref{thm:gaussZ}). We have also clarified what the restriction to the global orbit loses (section~\ref{sec:pullback}) and why the absence of a lower bound on the energy does not affect unitarity (section~\ref{sec:mass}).

The main remaining question is analytic. We have not shown that the image of the restriction map is dense, and a unitary realization requires quasi invariant measures on the orbit, after which the restriction to the Lorentz group should be a Mackey direct integral over double cosets. Density would follow from a smooth slice through $q$ transverse to the Lorentz orbit, which is a statement about Fréchet manifolds that the inverse function theorem does not provide. The perturbative methods of~\cite{Garbarz:2015lua} and the path integrals of~\cite{Maxfield:2026men,Mao:2026bms} suggest routes. A second question concerns observables beyond functions of the matrix norm, in particular heat kernels on the Lorentz group, whose amplitudes should be compared with the Wilson spool~\cite{Pannier:2025wil}, the Selberg zeta function approach~\cite{Bagchi:2023ilg} and the exact path integrals of~\cite{Maxfield:2026men}. A third concerns the symplectic structure of the singular near horizon currents of section~\ref{sec:softhair}.

The mechanism should extend to other theories, and we state what we expect. In anti de Sitter gravity the asymptotic symmetry group is a product of two Virasoro groups, and the winding geometries of~\cite{Garbarz:2014kaa} correspond to pairs of orbits, each of which may carry a cyclic label, $r_+$ and $r_-$. A rigid rotation acts on both sectors, and in the flat limit only the character of this diagonal rotation survives as a label of the superrotation stabilizer, so we expect $r$ to arise as the combination $r_++r_-$ of the labels of the diagonal rotation, while the other combination should become part of the continuous supertranslation data. This would give the cyclic label a holographic origin in the two chiral sectors. In higher spin extensions the developing map takes values in the flag variety of $SL(N,\R)$, and the stabilizer is again a centralizer of the lifted monodromy, now in the universal cover of $SL(N,\R)$. The centralizer of a regular hyperbolic element of $SL(N,\R)$ has $2^{N-1}$ components, so higher spin charges may enlarge the discrete factor rather than break it. In supersymmetric extensions the supercharges shift the weights by one, so a super BMS$_3$ multiplet built on $\Om_{m,\Delta}$ is expected to contain the sectors $r$ and $r\pm1$, of opposite statistics. Since the rotation through $2\pi$ acts on the supercharges by $\pm1$ in the Ramond and Neveu and Schwarz sectors, the boundary condition of the gravitino should constrain which pairs of cyclic labels form multiplets. Carrollian versions of BMS$_3$ offer further settings for the same mechanism.

Finally, the physical role of the winding sectors deserves study. Their orbits contain configurations of arbitrarily negative energy, so they do not describe stable particles. The conditional flat bundle construction yields a factor $e^{i\pi rk/m}$ for a loop of winding class $k$ when the connected stabilizer character is trivial. Whether such factors occur in a gravitational path integral or distinguish physical observables is not established by the present finite orbital functionals.

\section*{Data availability}
No data were created or analysed in this study.

\section*{Code availability}
The separately supplied reproducibility archive (version~1.0.5, matched to manuscript v20 and applicable to the unchanged formulae of this revision) contains the Python code used for the symbolic and numerical checks described in appendix~\ref{app:tech}, together with instructions, tested dependencies, numerical tolerances, a test runner and reference outputs. The final submitted supplement should include this archive with an integrity manifest. These computations provide supplementary verification and do not replace the analytical proofs.

%======================================================================
\appendix

\section{Finite cyclic sectors of \texorpdfstring{$B(2,1)$}{B(2,1)}}\label{app:B21}

This appendix explains how the finite cyclic little groups and the finite shift equation arise in $B(2,1)$~\cite{Melas:2017jzb,Melas:2021hil,Melas:2026iwe}. The group is $A\rtimes\SL$ with $A=L^2(P^1(\R),\lambda,\R)$, where $\upsilon=\cot(\theta/2)$ and $d\lambda=d\theta/2\pi$. With $\upsilon g=(\upsilon a+c)/(\upsilon b+d)$ and $\kappa_g(\upsilon)=((\upsilon a+c)^2+(\upsilon b+d)^2)/(\upsilon^2+1)$, supertranslations transform as $(T(g)\alpha)(\upsilon)=\kappa_g(\upsilon)\alpha(\upsilon g)$ and supermomenta as $(T'(g)\beta)(\upsilon)=\kappa_g(\upsilon)^{-2}\beta(\upsilon g)$. In the angle these are vector fields and quadratic densities, without Schwarzian term. A direct computation gives $k(\phi)\cdot\cot\frac\theta2=\cot\frac{\theta+2\phi}2$, so $k(\phi)$ acts by $\theta\mapsto\theta+2\phi$. This factor of two is the origin of the double cover structure: $-I=k(\pi)$ acts trivially with $\kappa_{-I}=1$, so it lies in every little group.

The little group of $\beta$ is $L_\beta=\{g:T'(g)\beta=\beta\}$. In the Hilbert topology every nonzero $\beta$ has compact little group, equal up to conjugacy to $SO(2)$ or to some $C_{2m}$~\cite{Melas:2021hil}. Connected noncompact little groups are excluded by an elementary argument. If the flow of $v\partial_\theta\in\sltwo$ fixes $\beta$, then $v\beta'+2v'\beta=0$, so $\beta=cv^{-2}$ on each interval where $v\neq0$. Hyperbolic and parabolic generators have zeros, near which $v^{-2}$ is not square integrable, so $\beta=0$, while an elliptic generator conjugate to $\partial_\theta$ gives the rotation invariant $\beta$ with little group $SO(2)$. Finite little groups are compact, hence conjugate into $SO(2)$, hence cyclic, and of even order because they contain $-I$.

A supermomentum is invariant under $C_{2m}=\langle k(\pi/m)\rangle$ if and only if $\beta(\theta+2\pi/m)=\beta(\theta)$, that is, if its Fourier modes $\beta_n$ vanish unless $m\mid n$. A standard representative is $\beta=\mu+\varepsilon\cos(m\theta)$ with $\varepsilon\neq0$ and $m\ge2$. The rotation $k(\phi)$ fixes it if and only if $\phi\in\frac\pi m\Z$, and since the little group is compact and contains the noncentral elliptic element $k(\pi/m)$, which fixes a unique point of $\Hb^2$, it is exactly $C_{2m}$. For $m\ge2$ the modes $n=\pm1$ vanish, so the translation projection is $(\mu,0,0)$, timelike for $\mu\neq0$.

The induced representation acts on functions with $\Psi(g\ell)=\sigma(\ell)^{-1}\Psi(g)$ for $\ell\in L_\beta$ by $(U(\alpha,g_0)\Psi)(g)=e^{i\langle T'(g)\beta,\alpha\rangle}\Psi(g_0^{-1}g)$. For $L_\beta=SO(2)$ and $\sigma_s(k(\omega))=e^{is\omega}$, the condition $\Psi(gk(\omega))=e^{-is\omega}\Psi(g)$ holds for all $\omega$, and differentiation gives the spin equation $(i\partial_\omega-s)\Psi=0$, which fixes the compact weight and, on the mass shell, gives the Pauli and Lubanski equation with helicity $s/2$. For $L_\beta=C_{2m}$ there is no one parameter subgroup in the little group and no differential equation. With $\sigma_r(k(\pi/m))=e^{i\pi r/m}$ the condition is the finite shift equation $\Psi(\xi,\omega+\frac\pi m)=e^{-i\pi r/m}\Psi(\xi,\omega)$, whose square integrable solutions are $\Psi=e^{-ir\omega}F$ with $F$ of period $\pi/m$, or $\sum_n\psi_n(\xi)e^{-i(r+2mn)\omega}$. The weights are fixed only modulo $2m$, and $F$ is a genuine degree of freedom on the fibre $K/C_{2m}$ of $\SL/C_{2m}\to\Hb^2$. For $\mu>0$ the Poincar\'e content is the ladder $r/2+mn$ with multiplicity one.

The geometry of a finite cyclic orbit is simple. The orbit is the three dimensional homogeneous space $\SL/C_{2m}$. When the translation projection $(\mu,0,0)$ is timelike, its stabilizer in the Lorentz group is $K$, and the orbit fibres over the two dimensional mass hyperboloid $\SL/K\cong\Hb^2$ with fibre the circle $K/C_{2m}$. The mass $\mu$ is therefore only part of the orbit data. Within the standard family $\mu+\varepsilon\cos(m\theta)$ at fixed $\mu$, the modulus $|\varepsilon|$ is an orbit invariant, because an element of $\SL$ preserving the rest vector $(\mu,0,0)$ lies in $K$, and a rotation can only change the phase of the $m$th Fourier mode, not its amplitude~\cite{Melas:2026iwe}. The Poincar\'e representation therefore does not determine the representation of $B(2,1)$, and the finite cyclic sectors carry information beyond the Poincar\'e quantum numbers.

We now spell out the cyclic equivariance and the conditions under which it produces a helicity ladder. Let $\chi_r$ be the character of $C_{2m}$ with $\chi_r(k(\pi/m))=e^{i\pi r/m}$, $r\in\Z/2m\Z$. A section of the induced bundle is a function on $\SL$ with
\[
\psi(g\gamma)=\chi_r(\gamma)^{-1}\psi(g),\qquad\gamma\in C_{2m},
\]
so the cyclic group acts from the right, on the argument of the wavefunction, and not on spacetime. Writing $g=s(\xi)k(\omega)$ and expanding in $\omega$, the condition for the generator $k(\pi/m)$ multiplies the coefficient of $e^{-is'\omega}$ by $e^{-i\pi s'/m}$, and comparison with $e^{-i\pi r/m}$ gives the Fourier congruence $s'\equiv r\pmod{2m}$. Conversely, every square integrable function whose Fourier modes in $\omega$ satisfy this congruence is a section. The physical rotation of a state through the angle $2\phi$ is the left translation by $k(\phi)$, which commutes with the right action of $C_{2m}$ and acts on a second set of weights, those of the left action of $K$. These left weights are not restricted to the class of $r$, only to its parity, because the central element $-I$ acts on both sides by $(-1)^r$. Confusing the two actions would identify the monodromy label with a rest frame angular momentum, which is incorrect. The right weights $s'$ become helicities only under the following conditions. The translation projection $(\mu,0,0)$ must be timelike, $\mu\neq0$, so that its Lorentz stabilizer is the compact group $K$ and each right weight space carries an irreducible massive representation of the Poincar\'e group by Wigner's construction. The helicity is then $s'/2$ because $k(\omega)$ rotates through $2\omega$, and the energy sheet is fixed by the sign of $\mu$. For $\mu=0$ the translations act trivially, the stabilizer of the projection is the whole Lorentz group, and the congruence describes only a representation of $\SL$, analysed in section~\ref{sec:harmonic}.

The case $m=2$ illustrates all features. For $\beta=\mu+\varepsilon\cos2\theta$ the little group is $C_4$ with $k(\pi/2)^2=-I$, its characters are $k(\pi/2)\mapsto i^r$, $r=0,\dots,3$, and the finite shift equation $\Psi(\xi,\omega+\frac\pi2)=i^{-r}\Psi(\xi,\omega)$ iterated four times gives periodicity in $2\pi$. The weights are $r+4n$ and the helicities $r/2+2n$, bosonic for $r=0,2$ and fermionic for $r=1,3$. Changing $\varepsilon$ or adding modes that are multiples of two changes the orbit but not the mass $\mu$. In the main text the little group $C_{2m}$ arises as the intersection of the stabilizer $\R\times C_{2m}$ of a point of $\Om_{m,\Delta}$ with the Lorentz group, and $\beta=2\pi(q+\kappa/2)$.

\section{Technical lemmas}\label{app:tech}

This appendix collects the identities and standard facts used in the text. The Schwarzian $\Sch(h)=(\log h')''-\frac12((\log h')')^2$ satisfies $\Sch(f\circ h)=(\Sch(f)\circ h)h'^2+\Sch(h)$, vanishes on Möbius maps, and $\Sch(f(\lambda\,\cdot))(x)=\lambda^2\Sch(f)(\lambda x)$. For $\widetilde\psi=(\psi\circ h)h'^{-1/2}$ one finds $\widetilde\psi'=(\psi'\circ h)h'^{1/2}-\frac12(\psi\circ h)h''h'^{-3/2}$ and
\[
\widetilde\psi''=(\psi''\circ h)h'^{3/2}-\tfrac12(\psi\circ h)h'^{-1/2}\Sch(h),
\]
because the two terms containing $(\psi'\circ h)h''$ cancel. This proves Lemma~\ref{lem:covariance}. From $\tan'''=2\sec^4+4\sec^2\tan^2$ one gets $\Sch(\tan)=2$. An element $B\in\SL$ acts on angles by $\tan\phi_B=\mu\circ\tan$ with $\mu$ Möbius, so the cocycle identity gives $2\phi'^2+\Sch(\phi)=\Sch(\tan)=2$, that is $\Sch(\phi)=2(1-\phi'^2)$ for every $\phi\in\tSL$. Elements of $G_0$ act on $\varphi=2\theta$ as $g(\varphi)=2\phi(\varphi/2)$, hence $\Sch(g)=\frac12(1-g'^2)$, which inserted in~\eqref{eq:coadfinite} gives~\eqref{eq:linearshift}. For two solutions of $\psi''=Q\psi$ the product $u=\psi_1\psi_2$ satisfies $u'''=2Q'u+4Qu'$, which for $Q=p/2\kappa$ is $\delta_up=0$.

The Sturm separation theorem~\cite{Hartman:2002ode} is used in Theorem~\ref{thm:nogo}. If $\psi_0,\psi$ are independent solutions, $W=\psi_0\psi'-\psi_0'\psi$ is a nonzero constant and zeros of $\psi$ are simple. At consecutive zeros $a<b$ of $\psi$, $\psi'(a)$ and $\psi'(b)$ have opposite signs, so $W=\psi_0(a)\psi'(a)=\psi_0(b)\psi'(b)$ forces $\psi_0$ to change sign in $(a,b)$.

For an increasing homeomorphism $C$ of $\R$ commuting with $\zeta$, the function $C^n-\mathrm{id}$ is periodic, $a_n=\sup(C^n-\mathrm{id})$ is subadditive, and $\tau(C)=\lim C^n(x)/n$ exists, is independent of $x$, is conjugation invariant and satisfies $\tau(\zeta^kC)=k\pi+\tau(C)$. If $C$ has a fixed point, $\tau(C)=0$, and if $\tau(C)\neq0$, $C$ has no fixed point.

The pullback statement used in Theorem~\ref{thm:restriction} is the following lemma.

\begin{lemma}\label{lem:pullback}
Let $H\subset G$ be closed, $\rho$ a representation of $H$, $G_0\subset G$ a subgroup and $H_0=G_0\cap H$. Then $G_0/H_0\to G/H$, $g_0H_0\mapsto g_0H$, is injective, and $[g_0,v]\mapsto[g_0,v]$ is a $G_0$ equivariant isomorphism from $G_0\times_{H_0}\C_{\rho|_{H_0}}$ onto the pullback of $G\times_H\C_\rho$.
\end{lemma}

\begin{proof}
If $g_0H=g_0'H$, then $g_0^{-1}g_0'\in G_0\cap H=H_0$, which gives injectivity. The map is well defined because $H_0\subset H$. If $(g_0,v)$ and $(g_0',v')$ are identified in $G\times_H\C_\rho$, then $g_0'=g_0h$ with $h\in H_0$ and $v'=\rho(h)^{-1}v$, which is the identification in $G_0\times_{H_0}\C$. Surjectivity on fibres is clear. This completes the proof.
\end{proof}

In terms of functions, restriction maps $H$ equivariant functions on $G$ to $H_0$ equivariant functions on $G_0$ and commutes with the left action of $G_0$. A homogeneous line bundle $\widetilde G\times_{\widetilde H}\C_\chi$ with $\chi$ trivial on $\widetilde H^0$ is associated with the covering $\widetilde G/\widetilde H^0\to\widetilde G/\widetilde H$ with deck group $\pi_0(\widetilde H)$, so it carries a canonical invariant flat connection, whose holonomy along the loop defined by an element of $\pi_0(\widetilde H)$ is the value of $\chi$ on that element, or its inverse for the opposite orientation. The connected covers of $G$ are $G^{(N)}=\widetilde G/\langle T^N\rangle$, in which the lift $\widetilde R_m$ has order $mN$.

For a unitary representation $V$ of $C_N=\langle\gamma\rangle$ and $P_r=\frac1N\sum_\ell e^{2\pi ir\ell/N}V(\gamma^\ell)$, the orthogonality $\sum_{\ell=0}^{N-1}e^{2\pi is\ell/N}=N\delta_{s,0}$ gives $P_rP_{r'}=\delta_{rr'}P_r$, $\sum_rP_r=1$ and $V(\gamma)P_r=e^{-2\pi ir/N}P_r$, and $P_r^*=P_r$. With $N=2m$, $\gamma^m$ acts by $(-1)^r$. The same computation shows that $f\in L^2(\R/2\pi\Z)$ with $f(\omega+\frac{2\pi}N)=e^{-2\pi ir/N}f(\omega)$ has Fourier support in $r+N\Z$, with the convention $f=\sum_sc_se^{-is\omega}$ of Theorem~\ref{thm:floquet}. If $w(\ell)$ depends only on $n\ell$, it is constant on cosets of the kernel of $\ell\mapsto n\ell$, of order $d=\gcd(n,N)$, and summing over the kernel gives the factor $\sum_{k=0}^{d-1}e^{2\pi irk/d}=d\,\delta_{d\mid r}$.

The integrals of section~\ref{sec:mass} follow from $\int_0^{2\pi}\frac{dx}{b+a\sin x}=\frac{2\pi}{\sqrt{b^2-a^2}}$ for $b\ge1>|a|$, obtained by residues. Differentiation in $b$ at $b=1$ gives $\int_0^{2\pi}\frac{dx}{(1+a\sin x)^2}=2\pi C^3$, the identity $\frac{\sin x}{1+a\sin x}=\frac1a(1-\frac1{1+a\sin x})$ gives $\frac{2\pi}a(1-C)$, and differentiation in $a$ gives $\int_0^{2\pi}\frac{\sin x\,dx}{(1+a\sin x)^2}=-2\pi aC^3$.

The proofs are given in the text. The computations described in the rest of this appendix are supplementary checks of specific identities at the stated parameter values, and they do not replace the proofs. Physical interpretations, such as superselection, saddle contributions or the role of the winding sectors in quantum gravity, are conjectural and are not checked. Statements of numerical agreement in this paper mean that the formula was checked numerically at the reported parameter values. Such checks concern specific values of $m$, angles, profiles and tolerances, and they do not establish identities for all values. All checks are contained in version~1.0.5 of the reproducibility package distributed with this paper as supplementary material. Each check is a function with an explicit tolerance, and a test runner reports a pass or a failure for each of them and exits with a nonzero status if one fails. The package contains a report of a complete run in a fresh virtual environment with the locked versions Python~3.13.17, NumPy~2.5.3, SciPy~1.18.1 and SymPy~1.14.0, on a Linux machine with two virtual Intel Xeon processors at $2.1$\,GHz. In that run all 255 checks passed, with a wall clock time of $22$ seconds. Runtimes on other machines may differ, and the package also allows each group of checks to be run separately.

The checks correspond to the results as follows. The normalization of the Haar measure is fixed analytically by Lemma~\ref{lem:haar}, and the checks confirm it in three independent ways: the determinant in its proof is evaluated exactly, the Jacobian of the parametrization is computed from finite differences of $g^{-1}dg$, and the two preimages of the parametrization are verified on two hundred matrices drawn with a fixed seed. The Iwasawa density $\frac12e^{-s}$ is checked in the same two ways, and the orbital integrals are recomputed in Iwasawa coordinates for $\vartheta=\pi/2,\pi/3$ and $\phi_0^F$, $\phi_1^F$ with the Gaussian profile, which tests the prefactor independently of the coordinates of Lemma~\ref{lem:haar}, with agreement within $10^{-11}$. The orbital integrals of Proposition~\ref{prop:gaussW} are evaluated by direct quadrature of the group integrand $\phi_i^F(xk(\vartheta)x^{-1})$ over $(\alpha,t)$ and, separately, by the closed forms, for $m=2,3$, all $\ell\notin\{0,m\}$, the three test functions and the profiles $e^{-0.6x}$, $(1+x)^{-3}$ and $x^2e^{-x}$, with relative tolerance $5\cdot10^{-10}$ and a comparison of two quadrature tolerances. The amplitudes of Theorem~\ref{thm:gaussZ} are formed from the directly integrated values for $m=2,3,4$ and all $r$ and compared with the theorem with tolerance $10^{-9}$. The same check verifies $\sum_rZ^\flat_r=0$ and the positivity of $\Delta(0,m;\phi_0^F)$ required by Corollary~\ref{cor:separation}, which for $m=2$ takes the values $1.067$, $0.1139$ and $2.835$ for the three profiles. The trigonometric sums in the proof of Theorem~\ref{thm:gaussZ} are checked for $2\le m\le14$. The radical form of $C_c$ in Proposition~\ref{prop:sl2proj} is compared with a thirty digit root of~\eqref{eq:etac}, and the resolvent cubic and the causal polynomial are verified exactly. The trace of the monodromy in Proposition~\ref{prop:floquet} is computed by integrating the Hill equation for four pairs $(m,\eta)$, with relative tolerance $10^{-9}$ and a convergence test, and equals $-135.81067$ for $m=3$ and $\eta=\frac12$. For Lemma~\ref{lem:unbounded} the identity $\int g\cdot p=E[w]$ is checked by three different evaluations for $f(y)=y+0.3\sin y$, the lowest Dirichlet eigenvalue of $-2\kappa\partial^2+p_{1,a}$ with $\kappa=1$ on intervals of length $5.6$ is negative for $\eta=0.4,1.8,3.0$ and stable under doubling of the grid, and $E[w_\lambda]$ is below $-10^3$ at $\lambda=1000$. The remaining checks concern equation~\eqref{eq:rep}, Propositions~\ref{prop:family}(i), \ref{prop:massrep} and~\ref{prop:singular}, the components of $P(p_{1,a})$, the thresholds $\eta_*(m)$, the selection rules and the symbolic identities of sections~\ref{sec:shift}, \ref{sec:bulk} and~\ref{sec:orbital}. They are listed with their methods and tolerances in the manifest of the package.
%======================================================================

\newcommand{\doi}[2]{\href{https://doi.org/#1}{#2}}
\newcommand{\arxiv}[1]{[\href{https://arxiv.org/abs/#1}{\texttt{arXiv:#1}}]}
\newcommand{\arxivold}[1]{[\href{https://arxiv.org/abs/#1}{\texttt{#1}}]}

\end{document}